\documentclass[onecolumn]{article}

\usepackage{arxiv}
\usepackage[numbers]{natbib}
\usepackage{subcaption}
\usepackage{graphicx}
\usepackage[hidelinks]{hyperref} 
\usepackage{url}
\usepackage{placeins}
\usepackage{multirow}
\usepackage{amsmath}
\usepackage{lipsum}

\newcommand\blfootnote[1]{%
	\begingroup
	\renewcommand\thefootnote{}\footnote{#1}%
	\addtocounter{footnote}{-1}%
	\endgroup
}

\newtheorem{definition}{Definition}
\begin{document}
\title{A Systematic Survey on Multi-relational Community Detection}

\author{
	Zahra Roozbahani\footnote{}\\
	Department of computer Engineering and IT\\
	University of Qom, Qom, Iran\\
	\texttt{z.roozbahani@stu.qom.ac.ir} \\
	\And
		Hanif Emamgholizadeh\\
	Department of computer Science\\
	Yazd University, Yazd, Iran\\
	\texttt{h.emamgholizadeh@gmail.com}
	\And
	Jalal Rezaeenour\\
	Department of Industrial Engineering\\
	University of Qom, Qom, Iran\\
	\texttt{j.rezaee@qom.ac.ir} \\
	\And	
	Mahshid Hajialikhani\\
	Department of computer Engineering\\	
	Azad University, Qazvin, Iran\\
	\texttt{m.hajialikhani@gmail.com} \\
}

%

\maketitle
\begin{abstract}
Complex networks contain various interactions among similar or different entities. These kinds of networks are called multi-relational networks in which each layer corresponds to a special type of interaction. Multi-relational networks are a particular type of multilayer networks in which nodes are similar entities; however, edges or communications demonstrate different types of interactions among similar entities.\\
In this survey, we study community detection methods for  multi-relational networks. The considered models are divided into two main groups, namely, direct methods and indirect methods. We put indirect methods in two classes: flattening and ensembling, and the direct methods are further divided into four groups which are: probabilistic methods, algebraic methods, modular-based methods, and graph feature-based methods. For each approach and each method, we explain their pros and cons.\\
Additionally, all the used datasets, in the multilayer community detection studies, are categorized into synthetic and real data. We elaborate on the most important datasets. Afterward, the utilized evaluation metrics by the papers are described. Finally, the current models' challenges and shortcomings are discussed. Finally, some suggestions for future research are developed. Putting all this together, this study, to the best of our knowledge, is the most comprehensive survey dedicated to multi-relational networks community detection.
\end{abstract}

\noindent \textbf{Keywords.} Network Science, Community Detection, Multi-relational Network, Multi-dimensional Network, Multiplex Network \blfootnote{$^*$ Corresponding author.} 

\section{Introduction}
Networks that describe real-life interactions contain different relations among the different entities. Multilayer networks show these various interactions. In other words, multilayer networks are the networks that contain several layers, and each of these layers consists of an independent single-layer network (traditional networks), each node can appear or disappear in each of these layers. Edges in each layer of a multilayer network have the same type, but they are different from other edges in other layers. Kivel{\"a} et al. \cite{kivela2014multilayer} provided a comprehensive study for introducing multilayer networks as well as related topics. In this study, various problems related to multilayer networks, such as network measures, coloring, community detection, etc. have been discussed. Mutlilayer networks are beyond the scope of this study, in which we only study multi-relational community detection. Multi-Relational Networks are the networks which have the same entities in different layers, whereas, the relations in each layer are from different nature. Among different topics related to multi-relational networks, one of the most challenging cases is community detection in multilayer networks. Before proceeding, it is worth mentioning that, in this study, we use "single-layer network" and "uni-layer network" interchangeably. Additionally, although a bit different, we also use "graph" and "network" in the same meaning in this paper.\\

The traditional perspective toward community detection was considering networks as single-layer graphs and aimed at uncovering the community structure of these uni-layer networks. There are a number of surveys dedicated to uni-layer community detection methods \cite{malliaros2013clustering,fortunato2010community,plantie2013survey}. Fortunato \cite{fortunato2010community} divided proposed methods for uni-layer community detection methods to four groups, namely, traditional methods, divisive algorithms, modularity-based methods, and spectral algorithms. Traditional methods utilize graph theory tools for detecting dense structure of the network \cite{pothen1997graph,kernighan1970efficient,barnes1982algorithm,radicchi2004defining}. Divisive methods aim at detecting inter-community edges and reveal community structure by eliminating the incident nodes to these edges \cite{girvan2002community,newman2004finding}. Optimization methods intend to detect a community structure which optimizes a quality measure of community structure \cite{newman2004fast,clauset2004finding,danon2006effect}. Eventually, spectral algorithms exploit eigenvectors and eigenvalues to extract community structures of networks \cite{eriksen2003modularity,simonsen2004diffusion,slanina2005referee,mitrovic2009spectral}. Fully consideration of these algorithms is beyond the scope of this study and interested readers are referred to \cite{fortunato2010community}.\\

Community detection in complex networks has been a challenging task. This problem grows when we take multilayer networks and multi-relational networks into account. Community detection in a network with only one layer is defined as the clustering of a network nodes into some groups in which the density of communication inside these groups is much higher than the density of their communications between the groups. These groups can be mutually exclusive, i.e., each node only belongs to a specific group. Or the groups can share some of their nodes with each other; in this case, the communities have some overlap with one another. Although clear for single-layer networks, due to having different types of relations, the definition of community density is not clear for multi-relational networks. Despite the significant amount of efforts dedicated to community detection in multi-relational networks, this field is still in its infancy. In the past few years, a considerable number of studies have been allocated to community detection for multi-relational  networks; however, there is not an agreed definition for community structure in multi-relational networks yet \cite{tagarelli2017ensemble}.\\

In this research, we study proposed methods for community detection in multi-relational  networks and discuss the strength and weaknesses of the methods and also current challenges.  To do so, we search in google scholar\footnote{https://scholar.google.com/}, DBLP\footnote{https://dblp.org/}, Scopus\footnote{https://www.scopus.com/}, ScienceDirect\footnote{https://www.ScienceDirect.com} with these keywords: multi-relational community detection, multi-dimensional community detection, multi-relational network clustering, multi-dimensional network clustering, and multiplex network clustering. We extracted more than one hundred papers. Among the extracted papers only, 40 papers directly address the main question of this study. In other words, these papers aimed at solving community detection in multi-relational networks' problems. The remaining papers were about data clustering. To put it differently, they intended to solve clustering problems in multi-dimensional datasets, regardless of the pair entities' communications.\\

The simplest way to deal with multi-relational networks is applying single-layer community detection methods on each dimentsion of network, which is the reduced form of a multi-relational network, or employing a single-layer network on each dimension separately and then combining these community structures. Indirect methods for  multi-relational community detection are divided into two main groups, namely, flattening and ensembling. These methods work on single-layer networks. Flattening approach simplifies multi-relational networks into a single-layer network and employs traditional single-layer community detection methods on this oversimplified network \cite{rocklin2013clustering,chen2017multilayer,dong2013clustering}. In contrast, ensembling methods deploy traditional single-layer community detection methods on each dimension of multi-relational network separately. Then, the method finds a consensus on detected communities in each dimension \cite{wilson2017community,didier2015identifying,cai2005community,berlingerio2013abacus,tagarelli2017ensemble}. Simplifying a multi-relational network to a single-layer network, or considering a multi-relational network as several single-layer networks mean ignoring invaluable information in the network.\\

For making the most of the presented information in multi-relational networks, recently, researchers have started to propose direct methods for community detection, which aims at finding communities of a multi-relational network concerning all the provided information by the network \cite{ali2019latent,interdonato2017local,cheng2013flexible}. Direct methods intend to extract communities from the entire structure of multi-relational networks by optimizing an objective function. A group of the proposed methods tries to extract the community structure of multi-relational networks utilizing a particular quality measure optimization. Almost all of the methods which lay on this group exploit a new definition of modularity measure \cite{newman2004finding} as all or part of their objective function. A framework of network quality functions was proposed by Mucha et al. \cite{mucha2010community}, which makes the study of arbitrary multislice networks possible. The authors' proposed method considers a multiplex method as an uncorrelated indirect single-network. Mucha et al. proposed a new modularity measure, which combines modularity measures defined in \cite{newman2004finding} with k-partite modularity presented in \cite{murata2010new,murata2010detecting}. Algebraic methods have also been utilized for detecting communities in multi-relational networks \cite{tang2009clustering}.

Another group of direct methods utilizes algebraic tools for detecting communities. These approaches try to extend non-negative matrix factorization or spectral methods for detecting communities in multi-relational networks \cite{ma2018community,dong2012clustering}.\\

Probabilistic methods are the next group of approaches that aim at finding the community structure of a multi-relational network using probabilistic methods. Stochastic Block mode \cite{ali2019latent} and Random Walk \cite{li2018community} are the tools which have been employed by probabilistic multi-relational community detection methods.\\

Having proposed a new community detection algorithm for multi-relational networks, the next step for the researchers is evaluating the proposed model. A variety of measures have been exploited by the researchers for assessing the quality of their proposed method. We dedicate a separate section to introduce and explain different evaluation methods. Evaluation methods with respect to the used dataset for evaluation can be classified into two classes: evaluation methods for the dataset with predefined labels, known as similarity-based quality measures and evaluation methods for assessing the quality of detected communities, known as quality-based evaluation methods.\\ 

Another challenge for researchers is collecting or selecting a dataset for evaluating the proposed methods. For paving the way or researchers, we provided an almost comprehensive collection of utilized data in the literature. Furthermore, for convenience, we presented publicly available datasets' hyperlinks. Utilized datasets in the literature lay on two groups: synthetic data, which is made by the algorithms, and real-world dataset, which is collected from real-world processes.\\

\begin{figure*}
	\centering
	\includegraphics[width=0.5\textwidth]{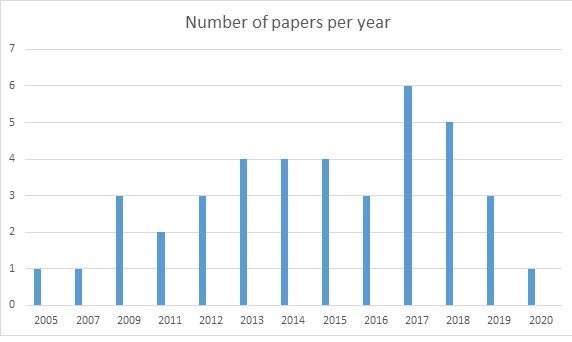}
	\caption{Histogram of studied papers.}
	\label{num_histo_pic}
\end{figure*}

Up to now, there is not a comprehensive study on the proposed methods for detecting community structures of multi-relational networks. Kim et al. \cite{kim2015community} survey has been the only work in this regard. In this study, seven papers for detecting community in two-dimension networks were considered. Moreover, five algorithms for detecting communities in multi-relational networks were introduced. It seems, there is an emergent need for categorizing the model because, in the past few years, a significant number of methods from a variety of viewpoints attempt to address multi-relational community detection problem. Also, challenges, opportunities, and potential should be discussed profoundly for shedding light on the future road of this research field. To overcome these shortcomings, in this study, we tried to consider all the multi-relational community detection methods deeply. We, additionally, discussed the challenges and potentials of this research topic. Figure \ref{num_histo_pic} denotes the histogram of considered papers in this study. As evident, more than half of the papers, which we study in this paper have been published after 2014, at which the previous survey on multi-relational networks was published.\\

The rest of this paper is organized as follows. First of all, in Sec. \ref{sec_def} we present some general definitions related to multi-relational networks. In Sec. \ref{sec_data}, we introduce used datasets by the proposed methods for evaluation. Then, in Sec. \ref{sec_methods}, we review the proposed methods for community detection in multi-relational networks. Next, in Sec. \ref{sec_eval}, we describe the evaluation methods for assessing the proposed methods, afterward, in Sec. \ref{sec_chall} we discuss existing challenges and future research. Finally, we conclude this paper in Sec. \ref{sec_concl}.

\section{Definitions}
\label{sec_def}
In this section, we define some basic concepts which will be utilized in the rest of this paper.
\begin{definition}
	\label{homo}
	\textbf{Homogeneous networks:} Network $G=(V,E)$, where $V$ is network vertices set and $E$ is network edges set is a homogeneous network if and only if the vertices and edges in the networks are from the same type \cite{sun2012mining}.
\end{definition}
\begin{definition}
	\label{hetero}
	\textbf{Heterogeneous networks:} Network $G=(V,E)$, where $V$ is the network vertices set and $E$ is the network edges set is a heterogeneous network if the vertices and edges in the networks are not necessarily from the same type. If $V_i$ is the set of nodes with type $i$, and $E_j$ is the set of edges with type $j$, then we have $V = \cup_i V_i$ and $E = \cup_i E_i$ \cite{sun2012mining}.
\end{definition}

\begin{figure*}
	\centering
	\includegraphics[width=0.5\textwidth]{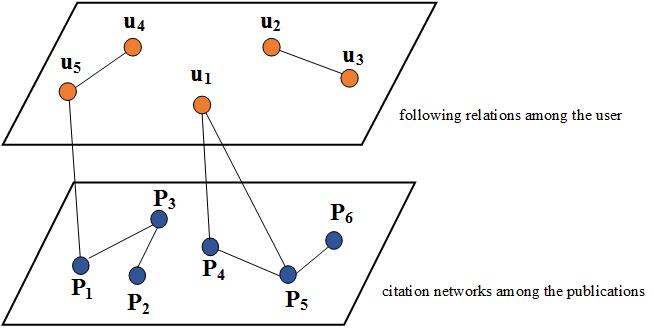}
	\caption{A multilayer network with two layers. A layer with users, as the layer nodes, and their following relations, as the layer edges. The next layer, with the publications, as nodes, and citation relations as edges. Finally, the inter-layer links indicate authorship relations.}
	\label{multilayer_pic}
\end{figure*}

\begin{figure*}
	\centering
	\includegraphics[width=0.5\textwidth]{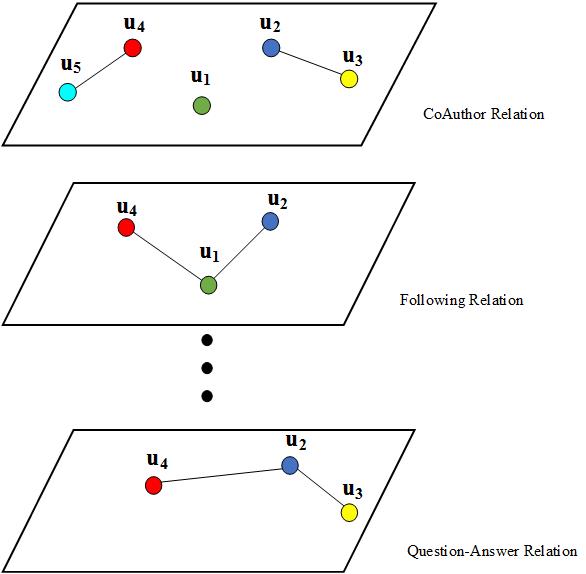}
	\caption{The figure denotes a multi-relational network with several layers. Nodes in all of the layers are users. Relations in all of the layers connect these users. In the figure, we have co-authorship, following, and question-answering connections.}
	\label{multirelation_pic}
\end{figure*}

\begin{definition}
	\label{multilayer}
	\textbf{Multilayer networks:} A multilayer network is made of some networks with their communications. Formally speaking, the network $G_M = (V_M, E_M)$ is a multilayer network where $V_M \subseteq V \times L$ is the set of nodes of the network. In this definition, $L$ is the layer set, and $V$ is the entire nodes of the network. Additionally, $E_M \subseteq V_M \times V_M$ is the set of the network edge \cite{kivela2014multilayer}. Fig \ref{multilayer_pic} indicates a multilayer network. In this multilayer network, the set of nodes contains users and publications. One of the layers in this multilayer network shows the following relations among the users. Another layer demonstrates citation networks among the publications. Finally, edges among these two layers' nodes indicate authorship relation between user and publication.
\end{definition}
\begin{definition}
	\label{multirelation}
	\textbf{Multi-relation networks:} A multi-relational graph contains one or more than one type of communications between the nodes. In other words, for a multi-relational network $G=(V,E)$, $V$ is the set of network's all nodes, which are the same among the layers. Additionally, if $1, \cdots, l$ are the relation's types in the network, we have $E = E_1 \cup E_2 \cup \cdots \cup E_l$, in which $E_i \subseteq V \times V$ \cite{paliouras2015discovery}. Figure \ref{multirelation_pic} shows a multi-relational networks. Nodes in all of the layers are researchers. Communications, in one of these layers, indicate Co-authorship relation; in the next layer, demonstrate the following relations, and finally, in the last layer, show question-answering relation. As stated, all of these relations are among the users. Multi-relation networks, also known as multiplex graphs; we use these two concepts interchangeably in this paper.
\end{definition}

\begin{definition}
	\label{aligned}
	\textbf{Aligned networks:} Two networks are aligned if some of they have some common nodes. If all the nodes are common in both of the networks the overall networks is called fully aligned \cite{zhang2013predicting}. A fully aligned network is also called \textbf{multi-relational network}.
\end{definition}
\begin{definition}
	\textbf{Community in multi-relation network: }A structure extracted from a multilayer network with some densely connected communities, which these communities are shared among all the layers, is called the community structure of multilayer networks \cite{kivela2014multilayer}.
\end{definition}
\section{Datasets}
\label{sec_data}
We introduce some of the important datasets utilized by the multilayer and multi-relational researchers for evaluating their proposed methods, and we elaborate on the most useful and important ones. We divided datasets into two main groups, that is, synthetic and real-world datasets. Synthetic networks are handmade networks in which the algorithm determines community structure. In other words, each node has a community membership label. The real-world datasets are extracted from the real world's networks. In the rest of this section, we explain the multilayer and multi-relational dataset in detail.

\subsection{Synthetic Networks}
Trying to simulate the signature of real-world networks, the algorithms of synthetic networks produce a graph with some layers and relations. The main advantage of these networks is their predetermined community structure in which all the nodes have a label corresponds to their belonging community. The disadvantage of these models is that the algorithms are able to simulate the real-world networks signature only partially \cite{hanif2016eval}.\\

\textbf{LFR Extension: } Pramanik et al. \cite{pramanik2017discovering} improved a well-known single-layer community detection benchmark, known as LFR, \cite{lancichinetti2008benchmark} to produce multilayer networks with predefined community structure. There are two parameters, that is, $\mu$ and $p$, where $\mu$ is used for determining nodes degree distribution, and $p$ indicates the noise level of the graph. The produced ground truth dataset can be used for assessing algorithms quality.\\

\textbf{mLFR: } Another extension to simple LFR \cite{lancichinetti2008benchmark} was proposed by Pizzuti et al. \cite{pizzuti2017many}. This method has two parameters. $\mu$ is mixing parameters and indicates the fraction of links shared by node with the nodes of its community. In other words, $\mu$ determine the rate of links which stay inside of community to the links which go out of community. The other parameter is DCC (Degree Change Chance) which control how much different the network layers are in terms of node degree. Higher DCC means more variation in distribution of a node in different layers.\\

\textbf{GN Extension: } GN benchmark \cite{girvan2002community} is another popular benchmark in single layer community detection literature. Ma et al. \cite{ma2018community} extended this famous algorithm by producing several networks as the layer of their multilayer networks. They create two kinds of multilayer networks. In the first one, the noise rate (number of edges that go out of the community) stays constant. However, in the next data set, each layer has its own noise rate.\\

\textbf{Albert-Barab{\'a}si based algorithm:} Liu et al. \cite{liu2018finding} proposed a synthetic dataset based on the famous Albert-Barab{\'a}si model \cite{albert2002statistical}. First, a set of subgraphs are made using Albert-Barab{\'a}si model. Then, a random connection/disconnection strategy for adding and removing edges among the subgraphs is employed to produce edges among the different subgraphs or layers.\\

\textbf{Structured Synthetic Random Multiplex (SSRM): }Structured Synthetic Random Multiplex (SSRM) was proposed by Loe et al. \cite{loe2015comparison}. The method starts with creating multiplex-communities that each one has high quality regarding a set of community quality measures. Afterward, the extracted community structure is modified to have a good quality for some of the quality measures, and lose its quality with respect to some other quality measures. The final step is combining these multiplex-communities into a single multiplex or multi-relational network community. This method was used by Pizzuti et al. in  \cite{pizzuti2017many}.\\

\textbf{Tang algorithm: } Tang et al. \cite{tang2009uncoverning} proposed a simple synthetic multilayer network that contains some network with a different number of nodes. Among the nodes, some relations with different types are produced. For each layer, community members connect with random chance. Additionally, a certain amount of noise by randomly connecting two different nodes is injected into the network. This method also was used by Amelio et al. in \cite{amelio2014cooperative}.\\

\textbf{English letters: } In this method, several point clouds corresponding to a set of English letters are formed in $R^2$ space. Each point clouds are made of some component of the Gaussian mixture model with different values for the mean and variance of Gaussian distribution. Each component represents a specific number of points with a particular color. An N-nearest neighbor graph is then established between each point in the cloud by labeling the weight of an edge between a pair of nodes as their Euclidean distance. If we use $l$ letter, by this algorithm, we will have $l$ layer network, with clusters corresponding to the color of points. This algorithm was used in \cite{dong2013clustering}.\\

\textbf{Uniform-Bernoulli edge distribution: }This algorithm was used by \cite{tang2012community} for producing multi-relational networks. For each layer, nodes are assigned to the pre-specified communities. Then, nodes inside the community are connected with uniform distribution to each other, and node pairs that are not in the same community are connected to each other using Bernoulli distribution. Finally, some noise is added between two randomly selected nodes.\\

\textbf{Matrix Blocking: } Inspired by \cite{condon1999algorithms}, Papalexakis et al. \cite{papalexakis2013more} introduced Matrix Blocking algorithm for creating multilayer networks with the prespecified community structure. In this algorithm, the network adjacency matrix is divided into a number of blocks, each indicator of a community. Each pair of nodes inside a block are connected using a user assigned probability. Finally, some noise is added to the whole network. This process is repeated for each layer of the network.\\

Synthetic datasets algorithm is summarized in Table \ref{tab_syn}. In \textit{Type} column, we indicated the type of produced network in the corresponding paper; however, for some of them, the extension for multilayer networks is easy to do. It is worth to mention that, most of these algorithms are just a simple extension of single-layer benchmarks. Whereas multilayer and multi-relational networks have unique characteristics, which putting a number of single-layer networks on top of each other does not necessarily produce them. Therefore, there is an immediate necessity for a standard synthetic benchmark to be used in multi-relational and multilayer community detection algorithms evaluation.
\begin{table*}[!]
	\begin{center}
		\caption{Table of synthetic datasets.}
		\label{tab_syn}
		\begin{tabular}{|c|c|c|c|}
			\hline
			Name&Used reference& Type &Link \\
			\hline
			\hline
			LFR Extension&\cite{pramanik2017discovering}& Multi-relation &-- \\
			\hline
			mLFR&\cite{pizzuti2017many}& Multi-relation &-- \\
			\hline
			GN Extension&\cite{ma2018community}& Multi-relation &-- \\
			\hline
			Albert-Barab{\'a}si based&\cite{liu2018finding}& Multi-relation &\href{http: //proj.ise.bgu. ac.il/sns/students.html}{Link\footnote{http: //proj.ise.bgu. ac.il/sns/students.html}} \\
			\hline
			SSRM&\cite{loe2015comparison}& Multi-relation &-- \\
			\hline
			Tang&\cite{tang2009uncoverning} \cite{amelio2014cooperative}& Multi-relational & --\\
			\hline
			English Letters&\cite{dong2013clustering}& Multi-relational & --\\
			\hline
			Uniform-Bernoulli edge distribution&\cite{tang2012community}  & Multi-relational &-- \\
			\hline
			Matrix Blocking&\cite{condon1999algorithms}& Multi-relation &-- \\
			\hline
		\end{tabular}
	\end{center}
\end{table*}
\subsection{Real-world dataset}
The algorithms have utilized a variety of real-world multilayer or multi-relational networks for assessing their qualities. There are a great number of collected datasets; thus, we only elaborate on the most important real-world datasets and present information of some of the other real-world datasets in Table \ref{tab_data_import}. Important datasets have been collected from 10 sources. In this section, we introduce these sources and corresponding datasets.\\

\textbf{DBLP: } DBLP is a bibliography website for computer science-related publications. In the dataset collected from this website, layers are different computer science conferences, publication relations, an so forth. This website is a popular source for collecting data. Four of the considered studies collected their dataset from this source \cite{,berlingerio2011finding,papalexakis2013more,hmimida2015community}. \\

Berlingerio et al. \cite{berlingerio2011finding} extracted published papers in DBLP. Layers correspond to the conferences held from 2010 to 2012. The final network includes 558800 nodes and 2668497 edges.\\

Hmimida et al. \cite{hmimida2015community} extracted a three-layer network from DBLP. The co-authorship layer includes 2809 nodes and 5109 edges; co-citation layer contains 2809 nodes and 36187 edges; eventually, co-venue layer consists of 2089 nodes and 251819 edges.\\

Interdonato et al. \cite{interdonato2017local}, using the DBLP website, created a 50-layer network. The layers are made of 50 well-known computer science conferences. The links in the layers are co-authorship relations. The network contains 83901 nodes and 159302 edges.\\

Papalexakis et al. \cite{papalexakis2013more} extracted two different sets of the networks, each of which corresponds to a different set of conferences. The networks include three layers. Nodes in each layer are the authors. One of the layers is made of the author to author citation relation. The other one is established by co-authorship relations. Finally, in the last layer, there is a link between two authors if they have at least three similar terms in the title or abstract of their papers.


\blfootnote{$^7$http: //proj.ise.bgu. ac.il/sns/students.html} 
\textbf{Internet Movie Database (IMDB): } Santra et al. \cite{santra2019structure} extracted their dataset from IMDB. The dataset contains three main layers. The first layer includes the actors. Two actors are connected if they have acted in the same move. The second layer is the director layer. In this layer, two directors have a relationship if they have direct at least one movie in the same genres. Finally, the movie layer consists of movies, and there is a link between two movies if their rating lay on the same range. In addition to this intra-layer relation, there also are three types of inter-layer relations: directs-actor, acts-in-movie, and directs-movies. The number of nodes in the actor layer is 4588, and the number of edges is 10207. The number of nodes in the director layer is 1091, and the number of edges is 91990. Finally, the number of nodes in the movie layer is 1000, and the number of edges is 78492 (the paper has provided the number of actors and density, for keeping consistency in our research, we extracted the number of edges using this two information). Another version of a multi-relational network extracted from IMDB was used in \cite{boden2012mining}, in which nodes are actors, link labels are the movie by which the link is made. There are four layers, namely, the first year of collaboration, the last year of collaboration, rental fees, and sold tickets. Overall this graph contains 300 vertices and 18368 edges.\\

\begin{table*}[!]
	\begin{center}
		\caption{Table of important real-world dataset.}
		\label{tab_data_import}
		\begin{tabular}{|c|c|c|c|c|c|}
			\hline
			Name&Used reference& \#Layers& \#Nodes& \#Edges &Link \\
			\hline
			\multirow{5}{*}{DBLP} &\cite{interdonato2017local}& 50& 83901& 159302 & \href{http://dblp.uni-trier.de/xml}{Link\footnote{http://dblp.uni-trier.de/xml}} \\ \cline{2-6}
			&\cite{berlingerio2013abacus}& 2536& 558800& 2668497 &NA \\ \cline{2-6}
			&\cite{hmimida2015community}& 3&\begin{tabular}{@{}c@{}}2809\\  2809\\  2809\\  \end{tabular} &\begin{tabular}{@{}c@{}}5109\\  36187\\  251819\\  \end{tabular} &NA \\ \hline
			\multirow{4}{*}{IMDB} &\cite{boden2012mining}& 4& 300& 18368 & NA \\ \cline{2-6}
			&\cite{santra2019structure}& 3& \begin{tabular}{@{}c@{}}4588\\  1091\\  1000\\  \end{tabular} &\begin{tabular}{@{}c@{}}10207\\  91990\\  78492\\  \end{tabular} &NA \\
			\hline
			\multirow{3}{*}{Biological} &\cite{didier2015identifying}& 4& \begin{tabular}{@{}c@{}}12110\\  2528\\  8839\\  91212 \\  \end{tabular} &\begin{tabular}{@{}c@{}}60669\\  36762\\  166761\\ 1105547 \\ \end{tabular} & NA \\  \cline{2-6}
			&\cite{paul2016null}& \begin{tabular}{@{}c@{}}Synapse\\  Iconic \end{tabular} & \begin{tabular}{@{}c@{}}253\\  2212 \end{tabular} & \begin{tabular}{@{}c@{}}453\\  2050 \end{tabular}  & NA \\ 
			\hline 
			Bibsonomy &\cite{hmimida2015community}& \begin{tabular}{@{}c@{}}User-Resource\\  User-Tag\\  Tag-Resource\\  Tag-User\\ Resource-Tag\\  Resource-User\\ \end{tabular} & \begin{tabular}{@{}c@{}}116 \\  116 \\  412 \\  412 \\ 361 \\  361 \\ \end{tabular} & \begin{tabular}{@{}c@{}}901 \\  985\\  2496\\  1956\\ 2814 \\  1685 \\ \end{tabular} & NA \\ 
			\hline 
			\multirow{3}{*}{arXiv} &\cite{wilson2017community}& 13& 14489& 59026 & \href{http://www.cs.cornell.edu/projects/kddcup
				/datasets.html}{Link\footnote{http://www.cs.cornell.edu/projects/kddcup
					/datasets.html}} \\ \cline{2-6}
			&\cite{boden2012mining}& 3& 13396 & 673800 & NA\\  \cline{2-6}
			& \cite{rocklin2013clustering}& 3& 30000 & NA &NA \\
			\hline
			AU-CS & \cite{ali2019latent,wilson2017community,interdonato2017local,liu2017improved}& 3& 6407 & 74862 & NA \\ 
			\hline 
			\multirow{4}{*}{Social Networks} & \cite{interdonato2017local}& 3& 6407 & 74862 & NA\\  \cline{2-6}
			& \cite{liu2018finding,ahmed2016multi}& 3& \begin{tabular}{@{}c@{}}183\\  39\\  106 \\  \end{tabular} &\begin{tabular}{@{}c@{}}240\\  23\\  97\\ \end{tabular} & NA\\ 
			\hline 
			Nokia &\cite{dong2013clustering,dong2012clustering}& 3& 136& 14042 & NA \\ 
			\hline 
		\end{tabular}
	\end{center}
\end{table*}

\textbf{Biological networks: } The biological networks are made of human gens or protein-protein relations. A multi-relational biological network has been extracted by Didier et al. \cite{didier2015identifying}. The dataset has been collected from different facets of cellular protein function, namely, (i) physical interaction (12110 nodes and 60669 edges), (ii) belonging of proteins to complexes (2528 nodes and 36762 edges), (iii) functional interactions extracted from pathway databases (8839 nodes and 166761 edges), and (iv) co-expression correlations derived from mRNA expression (91212 nodes and 1105547 edges). These four facets establish four interaction layer of the network.\\

\textbf{Bibsonomy: } Bibsonomy is a social bookmark and publication sharing system. Hmimida et al. \cite{hmimida2015community} have used Bibsonomy as the source of their extracted multi-relational networks. The dataset contains three networks. Each network includes two layers. The user network comprises two layers: User-based Resource (116 nodes and 901 edges) and User-based Tag (116 nodes and 985 edges). Tag network consists of two layers: Tag-based Resources (412 nodes and 2496 edges) and Tag-based Users (412 nodes and 1956 edges). Eventually, Resource network includes two layers: Resource-based tag (361 nodes and 2814 edges) and Resource-based User (361 nodes and 1685 edges).\\

\textbf{arXiv: } arXiv is a website for publishing pre-print versions of scientific articles. Boden et al. \cite{boden2012mining} used a multilayer network extracted from arXiv website. Three hundred keywords were used for extracting related papers. Each layer of this network contains publications related to a keyword, and relations are citation relations between the publications. This network includes 13396 nodes and 673800 edges. Wilson et al. \cite{wilson2017community} extracted a dataset from the arXiv website, which this data set was provided by De Domenico et al. \cite{de2015identifying}. The network is a multilayer collaboration arXiv network. Layers in this network are the scientific fields or sub-fields under which researchers collaborated. This network includes 13 layers, 14489 vertices, and 59026 edges. Rocklin et al. \cite{rocklin2013clustering} extracted 30000 high energy physics articles from ArXiv. Their network includes three layers Titles, Authors Abstracts, and Citations. Citation layer's relations are simple citation relations, and Title, and Author relations are made of the cosine similarity of the text using the bag of words model.\\

\blfootnote{$^8$http://dblp.uni-trier.de/xml} 
\blfootnote{$^9$http://www.cs.cornell.edu/projects/kddcup/datasets.html} 

\textbf{AU-CS: } This multilayer network was made of researchers' relation in a research institute in Australia. Layers are established based on different types of relations between users, such as Facebook.com relation, free time relation, work relation, collaboration relation, dining relation. This small networks includes 61 nodes, 620 edges, and 5 layers \cite{wilson2017community,interdonato2017local,ali2019latent}.\\

\textbf{Social networks: } Interdonato et al \cite{interdonato2017local} used a three layers. The nodes of these three layers are a set of users who have an account on YouTube, Friend Feed, and Twitter. Three layers correspond to three social networks. The overall network contains 6407 nodes and 74862 edges. \\

Another social network-based multilayer network, which was used by \cite{liu2018finding} and \cite{ahmed2016multi}, is \textit{Students' Cooperation Social Networks}. This network is built based on the student who participated in a course at Ben-Gurion University. The students were asked to log in to a specific website. Then using these students class relations a three-layers network including class partner relations, computer network relation (there is a link between two students if they have used the same computer for their assignments), finally, Time Network (students who have submitted their paper at the same time have time relation with each other.) was made.\\

\blfootnote{$^{10}$http://www.cs.umd.edu/hcil/VASTchange08/} 
\blfootnote{$^{11}$https://www.theialand.fr/en/product/spot-world-heritage} 
\blfootnote{$^{12}$http://lab41.github.io/surveycommunity-detection} 
\blfootnote{$^{13}$http://moreno.ss.uci.edu/ data.html} 
\blfootnote{$^{14}$http://deim.urv.cat/manlio.dedomenico/data.php} 
\blfootnote{$^{15}$http://www.gregsadetsky.com/aol-data} 

\textbf{Nokia Research Center: }This dataset was collected from the Nokia research center. It includes 136 cellphone users in Lake L'eman, Switzerland. The network includes three layers: GPS distance, Bluetooth scan, call communication. The final network includes 136 nodes, 14042 edges \cite{dong2013clustering}. \\

\textbf{C-elegant: }This weighted two-layer network is made of neurons, and the two types of neurons relations are chemical links or synapse and ionic channels. This network contains 253 nodes and 1695 edges in the synapse layer and 517 nodes in the ionic layer. This network was used in \cite{paul2016null}.\\

There are several real-world multilayer networks used for assessing the proposed algorithms; therefore, as mentioned before, we only elaborate on the most important datasets and summarize them in Tab. \ref{tab_data_import}. Some of the other datasets are introduced in Tab. \ref{tab_data_unimport}. 

\begin{table*}[!]
	\begin{center}
		\caption{Table of some of the real-world dataset which are not elaborated in the text.}
		\label{tab_data_unimport}
		\begin{tabular}{|c|c|c|c|c|c|}
			\hline
			Name&Used reference& \#Layers& \#Nodes& \#Edges &Link \\
			\hline
			Yelp&\cite{pramanik2017discovering}& \begin{tabular}{@{}c@{}}User\\  Location\\    \end{tabular} &\begin{tabular}{@{}c@{}}173697\\  13601\\  \end{tabular}& \begin{tabular}{@{}c@{}}244\\ 1627 \end{tabular}& NA \\
			\hline
			Meetup&\cite{pramanik2017discovering}&5727& 342773 & 31719 &NA \\
			\hline
			Cellphone networks&\cite{ma2018community}&10& 400 & 10500 &\href{http://www.cs.umd.edu/hcil/		VASTchange08}{Link\footnote{http://www.cs.umd.edu/hcil/			VASTchange08}} \\
			\hline
			Cancer networks&\cite{ma2018community}&4& 15054 & 1175250 &NA \\
			\hline
			Amazon co-shopping networks&\cite{ma2018community}&3& 12284 & 1570668 &NA \\
			\hline
			Airlines (EU-Air)&\cite{interdonato2017local}&37& 417 & 3588 &NA \\
			\hline
			BIOGRID &\cite{interdonato2017local}&7& 38936 & 342599 &NA \\
			\hline
			Reality Mining &\cite{interdonato2017local}&3& 88 & 355 &NA \\
			\hline
			Remote Sensing &\cite{interdonato2017local}&5& 642 & 4341 &\href{https://www.theialand.fr/en/product/spot-world-heritage}{Link\footnote{https://www.theialand.fr/en/product/spot-world-heritage}} \\
			\hline
			Facebook &\cite{alimadadi2019community}&NA& 500M & 2.1B &\href{http://lab41.github.io/surveycommunity-detection}{Link\footnote{http://lab41.github.io/surveycommunity-detection}} \\
			\hline
			Higgs-Twitter &\cite{tagarelli2017ensemble,paul2016null}&4& 456631 & 16070185 &NA \\
			\hline
			London Public Transportation &\cite{tagarelli2017ensemble}&3& 369 & 441 &NA \\
			\hline
			7th Graders &\cite{tagarelli2017ensemble,paul2016null}&3& 29 & 518 &NA \\ \hline
			Bank Employee's Network &\cite{amelio2014cooperative}&6& 16 & NA &\href{http://moreno.ss.uci.edu/ data.html}{Link\footnote{http://moreno.ss.uci.edu/ data.html}} \\
			\hline
			Kapferer Tailor Shop &\cite{liu2017improved}&4& 39 & 1018 &\href{http://deim.urv.cat/manlio.dedomenico/data.php}{Link\footnote{http://deim.urv.cat/manlio.dedomenico/data.php}} \\
			\hline
			YouTube &\cite{tang2009uncoverning,tang2012community,ahmed2016multi}&5& 15088 & 848003 &NA \\ 
			\hline
			QueryLog &\cite{berlingerio2013abacus}&5& 131268 & 2313224 &\href{http://www.gregsadetsky.com/aol-data}{Link\footnote{http://www.gregsadetsky.com/aol-data}} \\ \hline
		\end{tabular}
	\end{center}
\end{table*}
\section{Methods}
\label{sec_methods}
In this section, we discuss proposed community detection methods on multilayer or multi-relational networks. It is worth mentioning that when the researcher have used "multirelation" instead of "multilayer" in their papers we uses "multilayer". We divide the proposed methods into two main groups: direct and indirect methods. Indirect methods are divided into two subgroups, namely, flattening methods and ensembling methods. One of the first approaches for surmounting the challenges of multilayer networks was flattening \cite{rodriguez2010exposing}. This approach aims to reduce a multilayer or multi-relational network into a single-layer network and apply a single-layer community detection method for extracting network community structure. Although simple to implement, a significant amount of network information is lost in this approach. In fact, this approach does not exploit all capabilities of multilayer networks. Another indirect approach is ensembling in which, first of all, a single-layer community detection method is employed on each layer separately. Then, the best combination of these separately extracted communities is returned as the community structure of the multilayer network. Again, in this approach, some valuable information about the layers and their relations are ignored. Ensembling methods are different in the basic single-layer community detection and the combination of the detected community in each layer.\\

\begin{figure*}
	\centering
	\includegraphics[width=0.8\textwidth]{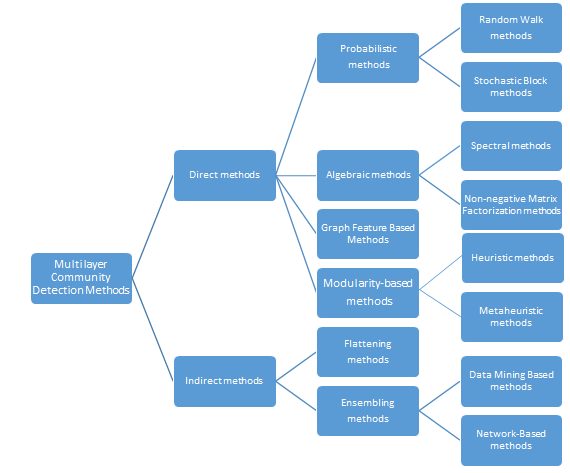}
	\caption{Multilayer community detection methods' chart.}
	\label{chart_pic}
\end{figure*}

Unlike indirect methods, direct methods are employed on multilayer networks directly. Often, direct methods are an extension of a single-layer community detection method. With respect to the basic idea, direct methods are categorized into four classes, that is, Modularity-based methods, Algebraic methods, Probabilistic methods, and graph-feature-based methods. Figure \ref{chart_pic} indicated the classification of multilayer community detection methods. In the rest of this section, we comprehensively elaborate on these classes and corresponding methods.\\

\subsection{Indirect Methods}
\label{indir_meth}
\begin{figure*}
	\centering
	\begin{subfigure}[t]{0.4\textwidth}
		\includegraphics[width=\textwidth]{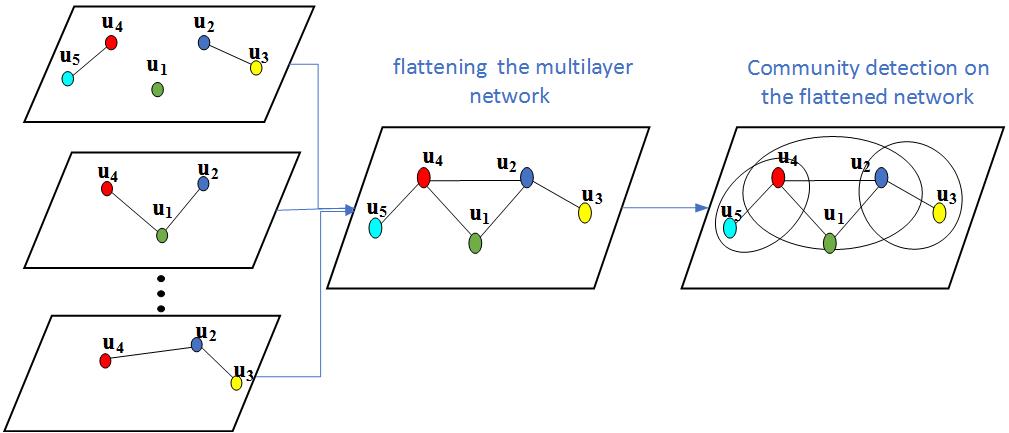}
		\caption{Architecture of flattening algorithms.}
		\label{flattening_pic}
	\end{subfigure}\hspace{15mm}
	\begin{subfigure}[t]{0.4\textwidth}
		\includegraphics[width=\textwidth]{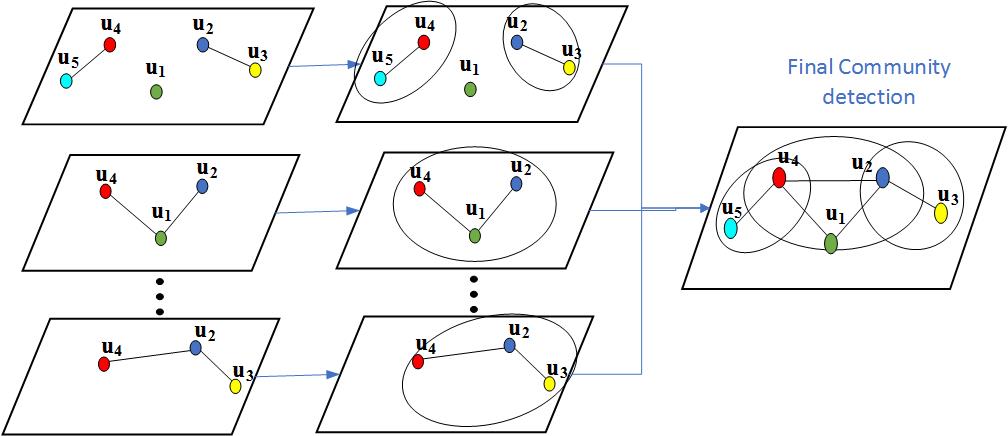}
		\caption{Architecture of ensembling algorithms.}
		\label{ensemb_pic}
	\end{subfigure}
	\caption{The left side figure indicates the architecture of a flattening algorithm for detecting communities in a multilayer network. The right side figure demonstrates the architecture of an ensembling algorithm for detecting communities in multilayer networks.}
	\label{indir_pic}
\end{figure*}

Indirect methods are the methods that use a single-layer community detection algorithm for extracting the community structure of multilayer networks. These methods either reduce a multilayer network to a single-layer network and then utilize a single-layer community detection methods on this final single-layer network (Figure \ref{flattening_pic}) or employ the single-layer community detection methods on each of the layers and then returns the best combination of this extracted community structure (Figure \ref{ensemb_pic}).  The former one is called flattening methods, and the latter one is called ensembling methods.

\subsubsection{Flattening}
The main aim of flattening methods is reducing a multilayer network to a single-layer network, and then deploying a single-layer community detection for extracting the community structure of the network. Therefore, these kinds of algorithms are indirect algorithms because they reduce multiple layers to a single layer and then use a single-layer community detection algorithm.\\

Algebraic methods also have been exploited by flattening algorithms. Dong et al. \cite{dong2013clustering} used an algebraic approach in their flattening method. The primary goal of this study was reducing of a multilayer network to a single-layer network to use the single-layer network as the input of a single-layer community detection method. The proposed algorithm, known as SC-ML (Spectral Clustering on Multi-Layer graphs), utilizes previous information for enhancing detected community structure's quality. This method maps the eigenvector of each layer of the multilayer network matrix on the space. Then, using Grassmann Manifolds, a combined mapping for multilayer networks with respect to each layer is produced. Finally, a clustering method is employed to detect the community structure of the multilayer network.\\

Rodriquez and Shirnavier \cite{rodriguez2010exposing} proposed another flattening method which applied algebraic methods. Their proposed method converts a multi-relational network into a uni-layer network, which is "semantically-rich." To do so, they modified a three-way representation of a multilayer network to a two-way tensor (adjacency matrix). This result was achieved by the algebraic path description method. Multi-relational path algebra is an algebraic structure that works on $n \times b$ adjacency matrix "slives" of an $n \times n \times m$ three-way tensor representation of a multi-relational network for producing a $n \times n$ path matrix, which represents "semantically-rich" single-relational network. Then, a single-layer community detection can be applied to this network for extracting community structure.\\

Two flattening methods were discussed in \cite{rocklin2013clustering}. The first one is an inverse method that uses a labeled data set. The labels indicated community membership. This method tries to achieve to an aggregation method to start from clusters and reach to a single layer network. As stated in this study, the final result of this inverse method depends on the quality of the single-layer community detection method. The second introduced method by the authors is a forward approach. This method aim at defining an optimization loop to find the parameters that achieve the most similar clustering to the ground-truth.\\

Some of the studies, inspired by information diffusion and label propagation or other similar approaches, try to map a multilayer network to a single-layer network and then detect the corresponding community structure.\\

The famous Label Propagation Algorithm (LPA) \cite{raghavan2007near,kianian2016semantic} was exploit as the community extraction stage of the flattening multilayer community detection algorithm \cite{alimadadi2019community}. In this method, the multilayer network, using similarity measures, is reduced to a single layer network, then LPA is employed. Finally, a postprocessing method is deployed to extract community structure. In the reduction stage, Jaccard, Common Neighbors and Adamic-Adar similarity measures were extended to consider the importance of each layer. Afterward, the extended LPA method was employed. In the proposed LPA, a function is used to evaluate the value of the labels; then, the most valuable label for this function is selected as the corresponding node's label.\\

This algorithm was proposed for exploring the best network based on friendship networks and activity networks, like, posting network, sharing network, endorsing (like) network. The study was evaluated on a directed network (Facebook); however, the authors have mentioned that this algorithm can be employed on an undirected network with adopting similarity measures. This method, like other LPA methods, is fast, and the number of clusters is not required in advance. However, similar to other LPA algorithms, the quality of the detected communities is not good enough. Additionally, the algorithm was tested on a small data set; therefore, more experimental studies are necessary to evaluate the quality of the method comprehensively.\\

For studying the features and characteristics of multilayer networks communities in the extracted community structure, Berlingerio et al. \cite{berlingerio2011finding} utilized a simple flattening method. The main objective of the author was to assess the attributes of the detected community for a multilayer network using a flattening algorithm and a single-layer community detection method, and they did not intend to introduce a new algorithm. They introduced two measures: \textit{Complementarity} and \textit{Redundancy}. \textit{Complementarity} was introduced to answer three questions: (i) how many of the layer exists in each of the communities, (ii) how many of the connected pair nodes are linked to each other in only one of the layers, and (iii) to what extent layers are distributed in the communities uniformly. \textit{Redundancy} is a measure to indicate the phenomenon for which a number of nodes that establish a community in a dimension also create a community in another dimension. Redundancy shows to what extent the existence of other dimensions in a multilayer community is meaningless. For extracting the community structure of a multilayer network, this method used some flattening methods; afterward, a single-layer community detection method was employed on the final single-layer network. The utilized flattening methods are: (i) adding an edge to a single-layer network between two nodes if there is a link between them in at least a layer of the network, (ii) adding weights to the edges concerning the number of layers in which corresponding link exists between two specific layers, and (iii) adding weights to the edge with respect to the number of common neighbors. This study provides an analysis of to what extent different layers collaborate in constituting a multilayer network community structure.\\

For assessing the impact of noise on the detected communities, Chen et al. utilized a flattening method \cite{chen2017multilayer,chen2016multilayer}. To do so, they implemented a convex layer aggregation method; then, using a spectral clustering algorithm, the community structure was extracted. The authors evaluate the resistance of the algorithm to the injected noise produced by the multilayer signal plus noise model. As the main contribution, this study was one of the first research, which considered noise impact on flattening multilayer community detection methods. 

\begin{table*}[!]
	\begin{center}
		\caption{Table Flattening methods.}
		\label{tab_flat}
		\begin{tabular}{|c|c|c|c|c|c|}
			\hline
			Reference&Algorithm Type& Weighted &Directed & Overlap & Evaluation Methods \\
			\hline
			\hline
			\cite{rocklin2013clustering}&Data mining& + &+ & - & Modularity \\
			\hline
			\cite{rodriguez2010exposing}&Algebraic& + & - & - & Statistical Analysis  \\
			\hline
			\cite{berlingerio2011finding}&Algebraic& - &- & - & Modularity \\
			\hline
			\cite{chen2016multilayer}&Algebraic& + & + & - & \begin{tabular}{@{}c@{}}Purity\\ NMI\\ Rand Index (RI) \end{tabular}  \\
			\hline
			\cite{dong2013clustering}&Algebraic& + & + & - & Accuracy  \\
			\hline
			\cite{alimadadi2019community}&Network& - & - & - & Homophily  \\
			\hline
		\end{tabular}
	\end{center}
\end{table*}

\subsubsection{Ensembling}
Some of the single-layer community detection methods can be combined to be used as a new algorithm for multilayer community detection. Or new measures and confinements can be assigned to the different kinds of networks, such as social networks or biological networks. These new measures and confinements can be utilized as the base of an ensembling method. The proposed ensembling algorithms lay on two main sets. In the first set, algorithms use network science tools to combine information about community structure in each layer. This information, then, is exploited for extracting the community structure of the multilayer network. The next set of algorithms use data mining and machine learning instruments for finding the best possible combination of the community structure of each layer.\\

\noindent \textbf{Network-Based methods}\\
As mentioned before, these kinds of methods utilize network science tools and network features for detecting the best possible combination of the community structure of each layer. Boden et al. \cite{boden2012mining} proposed an algorithm to be used on the homogeneous multilayer networks. The proposed algorithm aims to detect the best community structure, and this community structure should maximize the quality of detected communities in each layer. The authors used the \textit{quasi-clique} concept with the definition of a quality measure. Additionally, for reducing high node overlapping, the redundancy measure was taken into account.  In summary, the method provides communities with low overlap and high quality. The algorithm uses tree traverse on the network. For each node, the algorithm decides whether it is the member of a community or not, based on redundancy and quality measure. Paper's experiments showed better community quality and less run time for the proposed algorithm. This algorithm can be used for directed and weighted networks as well. \\

Boden et al. \cite{boden2012mining} does not take the topological community memberships of nodes into account. For extracting topological community memberships of nodes and also for preserving the multilayer topological information, Tagarelli et al. \cite{tagarelli2017ensemble} proposed a modularity based ensembling method for multilayer networks. In this method, firstly, the community structure of each layer is extracted. Then using a consensus method multilayer network's community structure is identified. Based on the authors' claim, this method preserves network topology. Therefore, the algorithm takes community membership amount as well as the type of edges among the nodes into account. This method can be used for multi-relational networks in which the nodes are the same in all layers. Based on the study experiments, the quality of this method is acceptable. However, its time complexity reduce its usability for big data.\\

Although \cite{boden2012mining} has proposed an ensembling method for multilayer networks community and \cite{tagarelli2017ensemble} has considered the topological community in its ensembling method, none of these studies did consider the importance of each layer into account. In the real world, the layer can have a different level of importance in a multilayer network. For instance, in a collaboration multilayer network, the layer, which includes the author collaboration relation, is much more critical than the layer, which contains financial supporters' relations (there is an edge between two supporters if they have supported the same project). Cai et al. \cite{cai2005community} addressed this problem. Instead of simply averaging on the layers, the authors exploited a heuristic method for calculating the importance of the layer with respect to users' needs. To this end, an objective function was introduced. The optimization of this objective function provides the importance of each layer. In this objective function, a new matrix based on user data is made. For creating this new matrix, communities of each layer of the network are extracted using a single-layer community detection method. Afterward, using these extracted communities, a labeled matrix is established for the multilayer networks. Eventually, by optimization of the objective function, the importance of each layer is determined.\\

Communities can overlap with each other. In other words, some nodes can be shared between two or more communities. For instance, a person can be a member of a sports team, and at the same time, a member of an art group. Wilson et al. \cite{wilson2017community} proposed an ensembling algorithm for overlapping community detection in multilayer networks. This algorithm extracts communities of each layer separately, and also address the problem of community detection in heterogeneous multilayer networks. Most of the proposed methods consider communities as mutually-exclusive sets; however, in the real-world people belong to more than an organization. Considering the heterogeneity of network in community detection methods broadens the horizon of the methods and enhance their ability to exploiting all the information provided by the networks. Although one of the main contributions of this study was detecting overlapping communities, it is not assessed comprehensively in the evaluation section. This algorithm can be used for community detection in weighted and attributed multilayer networks. The algorithm has high time complexity, which makes it unsuitable for big datasets. In contrast, this algorithm can be employed on a more significant range of data compared with other modularity-based community detection algorithms; additionally, it considers the structure of the layer in multilayer networks. This method is asymptotically consistent with multilayer stochastic block model and detects an overlapping community structure with high precision.\\

Network noise and its elimination have been a well-known challenge for researchers from the emergence of multilayer network science. Next two algorithms, using network science tools and heuristic methods, aim to overcome this long-lasting problem.\\

MLCDA is a multilayer community detection method, similar to the method proposed in \cite{wilson2017community}, extracts communities of each layer, and finally, provide the multilayer network's community structure using these extracted communities. This method was introduced in \cite{farzad2018multi}. It utilizes a hierarchical community detection methods on attributed graphs to find multilayer networks communities. The method contains two stages: synthesis and decomposition. The synthesis stage converts the primary network to a weighted network. In the decomposition phase, the community structure of each layer is extracted. This algorithm, in addition to noise reduction, achieves a reduction in time complexity of multilayer community detection; however, the algorithm's time complexity is still prohibitive for large data set. The algorithm can be utilized on directed, weighted, and attributed graph, and this is the main advantage of this algorithm.\\

Liu et al. \cite{liu2018finding} also aimed to address the overlapping community detection problem in multilayer networks. The identification of communities in multilayer networks requires a deep understanding of the dynamic and structure of layers. In real-world networks, there is an especial kind of relation among the nodes in a specific layer. Therefore, this study considered (i) the mutual impact of layers, (ii) the unique topological structure of each layer, and (iii) overlapping of communities into account, and thereby, proposed an algorithm for detecting community in a multilayer network. The algorithm calculates inter-layer and intra-layer similarity to the analysis of community structure. A dendrogram is created using these similarities, then, a new "community density" metric for cutting the dendrogram and extracting overlapping communities is utilized.\\  

The method works in this way. First of all, all layers of the network are combined, and a single-layer network is produced. Then, it extracts all link pairs and calculates the similarity of these link pairs. Next, these similarities are ordered in descending order. By mixing these similarities, a dendrogram is made. Finally, a newly introduced community density is calculated for each level of the dendrogram, and community structure with maximum density is returned as the final result. The returned community structure also includes overlapped communities. The main advantages of this method are the concentration on nodes interactions, reduction in the noise impact on the detected community structure of layers, and detection of smallest and densest community structure in real-world data.\\

For extracting the best combination of the extracted communities from each layer of heterogeneous multilayer networks Santra et al. \cite{santra2019structure} proposed an ensembling method. The method employs a decoupling approach for identifying the community structure of each layer in a heterogeneous network; then, it combines the detected communities using an optimization method. The study introduces a serial k-community for connected k layers, in which the community structure of the layers is combined in a particular order. The decoupling approach aims to detect the effectiveness of the extracted k-community structure for connected k layers. The algorithm includes these stages: (i) the communities of each layer using a single-layer community detection method are extracted. The detected community structure is called as 1-community. (ii) A bipartite graph with respect to the community structure of a couple of layers is made. In this bipartite graph, the nodes in each side correspond to a community in one of the communities (known as meta-nodes). The weighted edges of this bipartite graph indicate the relation between two communities in the selected layer. (iii) Afterward, the partial result of each layer (which are the meta nodes) are combined using a graph match as the composition function. This method is suitable for a two-layer network, which by employing it for multiple times, it can be extended to a k-layer network. This algorithm introduces several weight metrics that are customized for the community concept.\\

Finally, in \cite{didier2015identifying} Didier et al. demonstrated that in a heterogeneous and spars network with a high level of noise, like real biological networks, multiplex-modularity algorithm, as a special kind of ensembling method, outperforms other flattening and ensembling methods, which used in this study. Considering biological networks, the authors aimed at detecting communities using modularity. The algorithm utilized random graph models for simulating multiplex networks. This simulation is used for adding the topological features' information of the network to the algorithm. Three kinds of modularity measures, namely, aggregation modularity, consensus modularity, and multiplex modularity, were used in this paper. The study's experiments indicated that the multiplex modularity outperformed other methods. The algorithm can be used for a multi-relational weighted network in which all the nodes are shared in all the layers. The methods have been employed on biological networks, and more studies are needed for evaluating these methods on other types of networks.\\

\begin{table*}[!]
	\begin{center}
		\caption{Table of network-based ensembling multilayer community detection methods.}
		\label{tab_ensemb_net}
		\begin{tabular}{|c|c|c|c|c|c|}
			\hline
			Reference&Algorithm Type& Weighted &Directed & Overlap & Evaluation Methods \\
			\hline
			\hline
			\cite{boden2012mining}&cross-graph quasi-cliques& + &- & - & \begin{tabular}{@{}c@{}}GAMer-lg \\ GAMer-avg\\Cocain2\end{tabular} \\
			\hline
			\cite{tagarelli2017ensemble}&Modularity& + &+ & - & \begin{tabular}{@{}c@{}}Redundancy\\ Silhouette\end{tabular}  \\
			\hline
			\cite{cai2005community}&Graph Mining& - &- & - & Clustering Coefficient \\
			\hline
			\cite{wilson2017community}&Single-layer Community Detection& + &- & + & Precision  \\
			\hline
			\cite{didier2015identifying}&Modularity& + &- & + & Adjusted Rand Index \\
			\hline
			\cite{farzad2018multi}&Hierarchical& + &+ & - & \begin{tabular}{@{}c@{}}NMI\\ Precision\end{tabular}  \\
			\hline
			\cite{liu2018finding}&Hierarchical& + &- & +& \begin{tabular}{@{}c@{}}Accard Similarity\\ Recalrate\\Precision Rate\end{tabular}  \\
			\hline
			\cite{santra2019structure}&Max-Min Cut& + &- & +& \begin{tabular}{@{}c@{}}$w_e$\\ $w_d$\\$w_h$\end{tabular}  \\
			\hline
		\end{tabular}
	\end{center}
\end{table*}
\noindent \textbf{Data Mining Based Methods: }\\
Multi-View clustering is a well-studied topic in data mining, in which the methods intend to combine different views to a particular dataset. In other words, when there are several relations among a particular dataset's elements, and each of these relations is sufficient to learn from, multi-view clustering combines the final results to improve the clustering quality \cite{bickel2004multi,zhang2018binary,liu2013multi,chaudhuri2009multi}. It is worth noting that the relations in these kinds of datasets are not networks' edges necessarily. Therefore, they should be adopted to the network dataset for being used as a multilayer community detection algorithm. Data mining based methods which lay on the ensembling approach aim to detect the best combination of the community structures using data mining tools.\\

The algorithm introduced in \cite{berlingerio2013abacus} divides a multilayer network into several single-layer networks, and each of these layers' corresponding community structure is extracted. Afterward, for detecting the multilayer network's community structure, this single-layer community structure is combing using a frequent pattern mining method. In fact, the ability of this method in identifying a multilayer network's community structure depends on the basic single-layer community detection methods employed on each layer of the multilayer network. This algorithm works on the multilayer networks which have precisely the same nodes in all of the layers.\\

As mentioned above, this framework \cite{berlingerio2013abacus} starts with decomposing a multilayer network into several single-layer networks. Then, a single-layer community detection method is applied to these decomposed layers; the result is returned as a tuple $(dimension, community)$. Finally, a frequent pattern mining method is executed for detecting frequent community patterns in the layers.\\

This study proposed a flexible framework in which different algorithms and frequent pattern mining methods can be used to satisfy its user's expectations. However, a significant amount of information is lost during decomposition stage.\\

In \cite{ma2018community} the algorithm, first of all, factorize the network using a semi-NMF (Non-negative Matrix Factorization); then, the result of semi-NMF is used as the input of a spectral clustering method. Finally, a K-mean algorithm is employed on the modified version of the matrix, produced by spectral clustering, to extract community structure.\\

This algorithm starts with constructing prior information. To this end, a set of nodes are selected as seed nodes. To avoid selecting randomly, the algorithm chooses the nodes from a ranked list, which is produced using a ranking algorithm. The nodes with higher ranks are chosen as the seed nodes. A greedy algorithm is employed for extending detected dense graphs. Afterward, a vector is made to show the membership of each node to the detected subgraphs. Next, this vector is multiplied to itself, and the sum of the elements is calculated. The result is returned as a diagonal matrix in which the element $(i,j)$ indicates the probability that $i$ and $j$ lay on the same community. This diagonal matrix is added to the first matrix of the network. In the next stage, this final matrix is factorized by an NMF method. Then the factorized matrix is fed to a K-means clustering algorithm for detecting the final community structure. None of those mentioned above algorithms uses a semi-supervised algorithm; whereas, this algorithm adopts a semi-supervised approach. The quality of the detected community structure is increased in this method by injecting prior information to the algorithm. The algorithm can be used for dense and sparse networks, although its time complexity increase for dense networks. The network, to be used as input of this algorithm, should be unweighted and undirected. Also, all nodes should be the same in all layers of the network. Reducing the layers of a multilayer network in this method means ignoring precious information about the multilayer network. Additionally, the number of clusters should be determined in advance, and this issue is another confinement of the method.

\begin{figure*}
	\centering
	\includegraphics[width=0.8\textwidth]{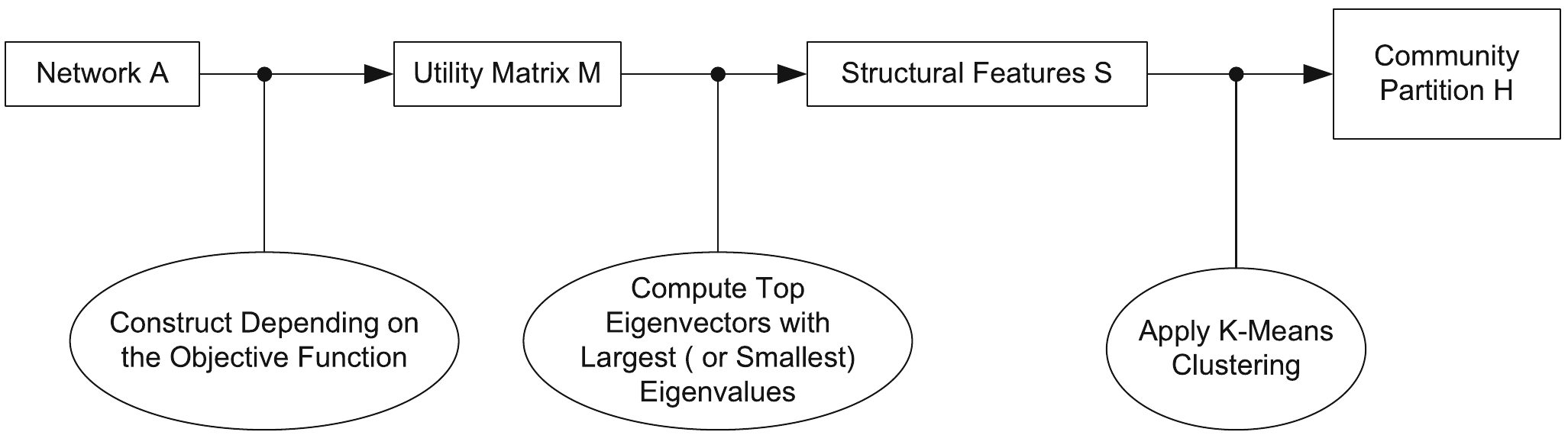}
	\caption{Four stages of the process of multilayer community detection \cite{tang2012community}.}
	\label{tang_pic}
\end{figure*}

Tang et al. \cite{tang2012community} study is, in fact, a review study on the flattening and ensembling methods. This study explains in which stage of a community detection process, we can apply each of the flattening or ensembling approaches. The authors explain a four stages framework for detecting the community structure of a multilayer network. For each of these four stages, a specific method for reducing a multilayer network to a single layer network was proposed. In the first stage, the multilayer network is represented as the matrix set. To convert this multilayer network to a single-layer network weight average of the relations was proposed. The matrix in the second stage is the utility matrix. The utility matrix can be constituted using the Laplace matrix, Modularity matrix, and so on. Weight average also can be used in this stage. The matrix in the third stage is made of structural features of the previous matrix. These structural features are eigenvectors and eigenvalues. Weight average is not practical in this stage because there is no unique answer concerning eigenvectors and eigenvalues. Therefore, for reducing the matrix of this stage, which is made of eigenvectors, Canonical Correlation Analysis (CCA), a standard statistical method, was proposed. The main aim of this method is to maximize pairwise correlation in the transition phase. The final stage is detecting the community structure of a multilayer network, which can easily be done using an ensembling method. Figure \ref{tang_pic} summarizes these four stages and the proposed algorithm in each stage. Based on the underlying algorithm, the proposed methods can be employed on directed and weighted graphs. The main shortcomings of the proposed algorithms, like other flattening and ensembling methods, are losing precious information in the reduction phase. Tab. \ref{tab_ensemb_data} summarize data mining based ensembling methods.

\begin{table*}[!]
	\begin{center}
		\caption{Table of data mining based ensembling multilayer community detection methods.}
		\label{tab_ensemb_data}
		\begin{tabular}{|c|c|c|c|c|c|}
			\hline
			Reference&Algorithm Type& Weighted &Directed & Overlap & Evaluation Methods \\
			\hline
			\hline
			\cite{berlingerio2013abacus}&Frequent pattern detection& + &- & + & \begin{tabular}{@{}c@{}}Conductance \\ Adamic similarity\end{tabular} \\
			\hline
			\cite{ma2018community}&Matrix Factorization& - &- & - & \begin{tabular}{@{}c@{}}NMI\\ Density\\ Association\end{tabular}  \\
			\hline
			\cite{tang2012community}\footnote{This algorithm could be inserted to all of the aforementioned table, i. e. \ref{tab_ensemb_net}, \ref{tab_flat}; however, we added it to this table as the summarization of flattening and ensembling methods.}&\begin{tabular}{@{}c@{}}Algebraic\\ Data mining\end{tabular} & + & + & - & \begin{tabular}{@{}c@{}}NMI\\ Modularity\\ CDNV \end{tabular}  \\
			\hline
		\end{tabular}
	\end{center}
\end{table*}

\subsection{Direct Methods}
A direct method implements a multilayer community detection method. It works on a multilayer network without considering it as several single-layer networks or reducing it into a uni-layer network. The direct algorithm can be divide into four main groups: Algebraic method, probabilistic methods, modularity-based methods, and graph-based methods. In the rest of this section, we elaborate on each of these groups.

\subsubsection{Algebraic Methods}

Algebraic methods utilize algebra for detecting the community structure of a multilayer network. Nonnegative Matrix-Factorization (NMF) and eigenvectors analysis are two popular approaches in this regard. Therefore, algebraic methods can be categorized into two sets of methods. In one of these sets, NMF is utilized, and in the other classes, spectral clustering and eigenvectors are exploited. The methods which use NMF factorize an adjacency matrix. This factorized matrix is further used to determine the membership probability of the node to the communities. Matrix dimension reduction methods lay on this set. Matrix dimension reduction methods aim to reduce the dimension of an adjacency matrix for eliminating noise and meaningless dimensions. To this end, often, a cost function and corresponding objective function are defined. By optimizing this objective function, the node-community membership matrix is returned. In spectral methods, most of the time, a transformed network using eigenvectors and eigenvalues are utilized for detecting community structure of a network.\\

\noindent \textbf{Nonnegative Matrix Factorization Methods}\\
One of the NMF methods is the Linked Matrix Factorization algorithm proposed in \cite{cheng2013flexible}. This algorithm intended to combine provided information in different layers in order to detect the common dense structure presented in all layers of the multilayer network. This study defines an objective function that its optimization reveals the common dense communities of layers. An impressive contribution of this method is its ability to combine non-structural information with structural information for enhancing the resulting quality. For instance, textual information can constitute a new network when the textual information relations provided by the network users are taken into account. This new network, in addition to the traditional structural network, is used as the input of the \textit{linked matrix factorization} method. The input of this network should be unweighted undirected networks, in which all the nodes in different layers are the same. However, the introduced method can easily be extended to work on weighted and directed graphs as well.\\

Some real-world networks are extremely dynamic and heterogeneous. However, most of the presented algorithms work on static and homogeneous networks. For overcoming this shortcoming, Lin et al. \cite{lin2009metafac} introduced an algorithm, which using hypergraph and tensor factorization, tries to detect the community structure of dynamic heterogeneous networks. As mentioned before, this method aims to define an objective function for factorizing the adjacency matrix. The optimization of this objective function ends up with the community structure of the network. Due to the high time complexity of optimization methods, most of the heuristic and metaheuristic methods look for the local optimums, and these local optimums can be far from the global optimum. This method has quadratic time complexity, which is a prohibitive element for this algorithm to be used on large datasets. This algorithm needs a pre-specified number of communities.  CGC (Co-regularized Graph Clustering) \cite{cheng2013flexible} is another algebraic multilayer community detection method which can be used on a wide variety of complex networks.\\

Cheng et al. intended to extend NMF to be used on multilayer networks \cite{cheng2013flexible}. To do so, they introduced an objective function for community detection in each layer. Then, two loss functions were defined to combine the layers relations with the objective function of the single-layer network. The first loss function tries to keep the number of communities in all the layers the same. However, using some mathematical methods, the second loss function release this constraint. Afterward, utilizing these loss functions, a new objective function for detecting communities of multilayer networks were introduced. Eventually, for optimizing this objective function, an iterative algorithm was utilized. The result of this final stage is the community structure of the multilayer network. The algorithm in the study uses many-to-many cross-domain instance relationship definitions as the required input network, and this definition makes the method flexible to be extended to a variety of multilayer networks. For instance, weights can be added to inter-layer edges, or hyperedges can be used in the network. Although the algorithm works on many-to-many cross-domain relations, it considers the same type of nodes in the layers. Therefore, this version of the algorithm is not applicable to heterogeneous networks.\\

\noindent \textbf{Spectral Methods}\\
For maximizing modularity, three methods were introduced in \cite{tang2009uncoverning}. First of all, the authors proposed two basic objective functions for optimizing the modularity of a multilayer network. The first objective function aims at maximizing the average of modularities for all layers. The second objective function intends to maximize the modularity of the multilayer network. In the first objective function, strong relations in a layer imposes more impact on the final result compared with the layer's weight in the multilayer network. However, in the second objective function, the weight of each layer is taken into account separately. The authors indicated that there is no guarantee about the ability of the second objective function for detecting efficient community structure. The third proposed method in this study is Principal modularity maximization (PMM). This method contains two stages. In the first stage, the method calculates the modularity matrix (new modularity measure), and $I$ are extracted as structural features of the modularity matrix. Then, the eigenvectors with positive eigenvalues (for eliminating noises) are selected and constitute a new matrix of $X$. Then SVD (Singular Value Decomposition) is deployed on $X$. Afterward, the K-means method is applied to the resulting decomposition of $X$ to extract the common community structure in the multilayer network. PMM can work on directed and weighted networks. The main drawback of this method, similar to other SVD-based methods, is the time complexity of SVD for big matrix, which makes this method impractical for big networks.\\

Using the Laplacian matrix of a network is one of the most popular methods for extracting community structure in the network because it increases the quality of final results \cite{fortunato2010community}. This concept was used in \cite{dong2012clustering} to extract the community structure of multilayer networks. This method is easy to understand and implement that these two advantages make it popular in the multilayer community detection field. In the single-layer Laplacian-based spectral community detection, the Laplacian matrix is calculated from the network (weight) matrix. Then, eigenvectors of the Laplacian matrix are computed, and $k$ biggest eigenvector is selected. These selected vectors are mapped on the space, and the K-means algorithm is employed to extract dense subspaces. This concept is extended in \cite{dong2012clustering} to be used on multilayer networks. In this extended version, the Laplacian matrix is calculated for each layer. Then an objective function, defined in this study, for all of the Laplacian matrices, is optimized. The result is matrix $P$, which is made of common eigenvectors of the multilayer network. In fact, $P$ includes joint eigenvectors of all layers. Eventually, like single-layer networks, the K-means algorithm is implemented on the $P$ matrix to extract the multilayer network's communities. This algorithm is practical for the weighted and directed multilayer networks, and the nodes in the layers should be the same.\\

In a recent study, Paul et al. \cite{paul2020spectral} studied the performance quality of aggregation, direct, and ensembling methods when the matrix which these methods work on is derived by the algebraic approach. Several baseline methods were utilized in this theoretical and comparative study. As the aggregation method, two spectral clustering methods \cite{ng2002spectral,rohe2011spectral} were applied on the mean of the adjacency matrix. Two algorithms were employed as direct methods, namely \cite{tang2009clustering}, which introduced above, and \cite{kumar2011co}, which is a multi-view clustering method. This multi-view clustering method's objective function has two elements. The first one is the usual association cut spectral clustering objective function for different layers, and the second one is the penalty function that tries to maximize cohesion of the eigenspace extracted from different layers. Finally, as an ensembling method, the authors employed a technique in which a spectral kernel was defined for each layer, then the aggregation of the kernels was calculated. This approach is similar to assign community structure for each node in each layer and then create a "module allegiance matrix" on each layer of the network. Finally, applying a consensus method \cite{braun2015dynamic}. The authors examined these methods on a wide range of multilayer networks in which the number of nodes, layers, and community increases. They conclude that for the sparse matrix, when each layer does not contain a significant amount of information, direct methods outperform both aggregation and ensembling methods.

Tab. \ref{tab_algeb} summarize the proposed algebraic methods for detecting community structure in multilayer networks.

\begin{table*}[!]
	\begin{center}
		\caption{Table of algebraic multilayer community detection methods.}
		\label{tab_algeb}
		\begin{tabular}{|c|c|c|c|c|c|}
			\hline
			Reference&Algorithm Type& Weighted &Directed & Overlap & Evaluation Methods \\
			\hline
			\hline
			\cite{cheng2013flexible}&Matrix Factorization& Inter-layer edges&- & - & \begin{tabular}{@{}c@{}}Accuracy\end{tabular} \\
			\hline
			\cite{lin2009metafac}&TMF& + &- & - & \begin{tabular}{@{}c@{}}P@10\\ NDCG\end{tabular}  \\
			\hline
			\cite{tang2009clustering}&Matrix Factorization& + & - & - & \begin{tabular}{@{}c@{}}NMI \end{tabular}  \\
			\hline
			\cite{dong2012clustering}&Spectral clustering& + & - & - & \begin{tabular}{@{}c@{}}NMI\\Purity\\RI \end{tabular}  \\
			\hline
			\cite{tang2009uncoverning}&Spectral clustering& + & + & - & \begin{tabular}{@{}c@{}}NMI\\ Cross-dimension network \\Validation\\ NMI \end{tabular}  \\
			\hline
		\end{tabular}
	\end{center}
\end{table*}

\subsubsection{Probabilistic Methods}
Based on the basic probabilistic approaches which have been utilized by community detection methods for multilayer networks, we divide probabilistic methods into two main categories: stochastic block methods and random walk methods. In the rest of this section, we explain these methods.\\

\noindent \textbf{ Stochastic Block methods}\\
Han et al. in \cite{han2015consistent} introduced a multi-graph stochastic block model in which the structure of the classes are the same. The main objective of the algorithm is to provide an accurate estimation of class memberships, using the stochastic block model. Spectral clustering, as the base of this method, provides promising results when it estimates the average and maximum similarity of a particular set of nodes for multilayer networks with more layers. The proposed method in \cite{han2015consistent} is suitable for the multilayer networks, which have a lot of layers or dimensions, and also nodes are not necessarily the same among the layers. Another advantage of this method is an increase in the result accuracy in comparison to pure stochastic methods. This algorithm also is robust when a block or an experiment result is eliminated. Indeed, even the algorithm loses a set of experiment results; it is able to reconstruct the results using statistical analysis. However, error freedom in this method is lower than the error freedom of a single-layer network. For the significant errors in the experiment, the algorithm ends up with a more substantial error. In other words, in the cases in which there is some lost data, the pure stochastic block model is outperformed by a single layer pure stochastic method. In the \cite{valles2016multilayer}, the authors proposed a probabilistic method to work on the networks in which the nodes in the layers are not necessarily the same. This method is more trustable in detecting fake and lost links compared to the algorithm introduced in \cite{han2015consistent}.\\

Valles et al. \cite{valles2016multilayer} suggested a Stochastic Block Model (SBM) for multilayer networks, which predict the structure of real-world networks. SBM is a generative model for stochastic networks, and these networks contain several communities with similar degree distribution. Multilayer SBM, as mentioned before, is an extension of single-layer SBM in which links in different layers are generated and collected by a variety of methods. This study deeply analysis a probabilistic approach for detecting optimized multilayer SBM. Due to being uncontrollable, an estimation is utilized for determining multilayer SBM. Stochastic block modeling is a popular method for classifying people based on their social relations.  In this study, this method is extended to be used on multilayer networks for detecting social-relation clusters. This method tries to assign an identity for each block; at the same time, unlike other methods, it takes heterogeneity of the network's node into account. The novel statistical analysis proposed in this study intends to identify rules of established social relations among the actors. This algorithm can be used on the networks in which each layer has a different number of nodes. Reduction in the error, increase in the accuracy of the system, making a balance between clusters, and implementation simplicity are the main advantages of this method. On the other hand, the time complexity of this method is high, which makes it impractical for the big networks. Huang et al. \cite{huang2018fusion} proposed a method for identifying communities in the multiplex network, using a probabilistic approach.\\

Huang et al. \cite{huang2018fusion} formulated the community detection method as a Bayesian inference problem with SBM parameters and proposed a belief propagation method for addressing this problem. This study introduced an asymptotically optimal algorithm for belief propagation, and the proposed algorithm was employed on the networks described by the stochastic block model. The nature of the belief propagation algorithm provides the opportunity of running the algorithm computation in parallel or on distributed computers. In summary, this study proposed a Bayesian model for multiplex graphs, which was solved using a belief propagation method. Ali et al. \cite{ali2019latent} suggested another probabilistic method which works on weighted networks.\\

Ali et al. \cite{ali2019latent} proposed a method for simultaneously detecting shared and unshared communities between heterogeneous weighted networks using Weighted Stochastic Block Model (WSBM). This algorithm is able to extract the community structure of every layer, like the unique community structure of a layer. The algorithm works on (weighted) multilayer networks. This algorithm is capable of detecting shared and unshared communities among the layers in weighted or unweighted multilayer networks. The number of communities is not needed in this method. Compared to algebraic methods, this algorithm has lower time complexity; however, it is not low enough to work on big networks. Additionally, the probabilistic base of this method makes inaccurate in detecting communities with low density.\\

The proposed method in \cite{lei2020consistent} is a probabilistic method that used a multilayer tensor instead of the adjacency matrix of the layers. This algorithm works on multi-dimensional undirected networks in which nodes are the same in different layers. Experimental dataset, used in this study, is small, and more evaluations are needed for comprehensive assessment of this model.\\

\noindent \textbf{Random Walk Methods}\\
A random walk commences its journey from a node of a multilayer network. In the next step, it can select staying in the current layer or going to another layer, with particular probability for each case, to follow its path. With this in mind, a set of nodes are considered as a community if the likelihood of staying inside this set for a random walk is more than going out. The primary motivation of using this approach is detecting community structure, which is good on all the layers, even if the extracted communities are not the best of every single layer. A random walk based method was proposed in \cite{zhou2007spectral}. This study defined an objective function which its optimization provided community structure. This objective function was introduced on two matrices, namely, the transition matrix and stationary matrix of a random walk. The proposed method is extended to be used for data classification. This classification version extracts the similarity of classification and clustering problems. The main advantage of this algorithm is its flexibility in working with weighted, unweighted, directed, and undirected networks.  \\

The proposed method by Li et al. \cite{li2018community} identifies the core nodes with respect to the repeatability of multi-layer nodes. Then, based on inter-layer trust, random walks within the layers and outside of the layers were established. Using a random walk, for each unclustered node, the algorithm determines the node membership to each of the clusters, and the algorithm adds the node to the clusters with the highest membership value. This method takes all the edges into account and rapidly adapts to the changes in the network. Additionally, this method is fast, concerning time complexity, and accurate when the input networks are social networks. The method does not work on weighted and directed networks. Table \ref{tab_prob} summarize probabilistic multilayer community detection methods.\\

\begin{table*}[!]
	\begin{center}
		\caption{Table of probabilistic multilayer community detection methods.}
		\label{tab_prob}
		\begin{tabular}{|c|c|c|c|c|c|}
			\hline
			Reference&Algorithm Type& Weighted &Directed & Overlap & Evaluation Methods \\
			\hline
			\hline
			\cite{ali2019latent}&WSBM& +&- & + & \begin{tabular}{@{}c@{}}NMI\\ Precision\end{tabular} \\
			\hline
			\cite{lei2020consistent}& SBM with tensor& -&- & - & \begin{tabular}{@{}c@{}}Statistical analysis\end{tabular}  \\
			\hline
			\cite{huang2018fusion}&SBM&- & - & + & \begin{tabular}{@{}c@{}}Statistical analysis\end{tabular}  \\
			\hline
			\cite{valles2016multilayer}&SBM& - & - & + & \begin{tabular}{@{}c@{}}AUC\\Fn\\FP\\Connectivity \end{tabular}  \\
			\hline
			\cite{han2015consistent}&SBM& - & - & - & \begin{tabular}{@{}c@{}}CEA \end{tabular}  \\
			\hline
			\cite{zhou2007spectral}&Random Walk& + & + & - & \begin{tabular}{@{}c@{}}Precision\\ Recall \end{tabular}  \\
			\hline
		\end{tabular}
	\end{center}
\end{table*}

\subsubsection{Modularity-Based Methods}
One of the popular approaches in detecting communities in single-layer networks is the optimization of an objective function that this objective function aims to maximize a well-known quality measure, that is, modularity. In the same line, objective functions and modularity measures have been used in some community detection methods for extracting multilayer networks community structure. The methods which optimize a modularity-based objective function are divide into two groups, namely, heuristic and metaheuristic methods. The heuristic methods propose a creative approach for optimizing an objective function. In contrast, metaheuristic methods utilize higher-level procedures design for estimating sufficiently good answers in comparison to the global optimum. In the rest of this section, we explain modularity based multilayer community detection methods.\\

\noindent \textbf{Heuristic Methods}\\
The first attempt for addressing multilayer networks community detection problem was made by Liu et al. \cite{liu2014framework}. The authors proposed composite modularity as a new quality measure. This measure works on heterogeneous multi-relational networks. The measure decomposes the heterogeneous multi-relational networks into several subnetworks and then integrates the modularity in each subgraph. Because the modularity optimization is an NP-hard problem, the authors proposed an intuitive algorithm for optimizing the objective function. Their intuitive algorithm utilize Louvain algorithm \cite{blondel2008fast} algorithm. To speed up traditional Louvain, the authors proposed the Louvain-C algorithm, which uses divide and rule strategy. This algorithm includes three steps:
\begin{enumerate}
	\item Detecting the community structure in each of the subgraphs, separately. Modularity and k-partite modularity optimization can be used in this step. When a node got involved in the different subgraphs, the node may assign to diverse communities. Thus, in this stage, the communities of the layers are detected.
	\item Combining the detected communities and deriving their constraints.
	\item Optimizing the composite modularity with respect to the constraints detected in step 2. In this step, considering the constraints, a new network is built, similar to the second stage of the Louvain algorithm \cite{blondel2008fast}. Nodes of this now network correspond to a group of nodes in the first network. The algorithm also recalculates the weights of nodes' links. By doing so, composite modularity runs on a much smaller network much rapidly and without losing the overall accuracy.
\end{enumerate}
This algorithm works on heterogeneous weighted multilayer networks. Due to its low time complexity, it is suitable for large networks. The evaluation of this study was limited to only one dataset; therefore, more experiments are necessary for having a broader horizon about the quality of this method. For extracting better communities from multilayer networks, better quality measures are required. For addressing this shortcoming Paul et al. \cite{paul2016null}, proposed several multilayer modularity measures and corresponding null model, motivated by empirical observations.\\

There are two sets of quality measures defined in \cite{paul2016null}: multilayer configuration model (MLCM) and the multilayer expected degree (MLED) model (these sets of measures are based on traditional modularity measure). In the next stage, these proposed modularity measures are optimized for detecting communities. These methods can be divided into two broad classes, namely, independent degree (ID) models and shared degree (SD) models. ID models include a free parameter in each of the layers for each node. In contrast, SD models share the node's information among the layers. SD contains fewer parameters, and therefore, they are more economical, which makes them suitable for sparse multilayer networks. The main objective and also advantage of this study is proposing some community detection methods which preserve information of different layers and integrate them for extracting better community structures.\\

Pramanik et al.'s method extended modularity based multilayer community detection to work on directed networks \cite{pramanik2017discovering}. The primary goal of this study was to propose a modularity measure to work on multilayer networks directly. Indeed, the authors aimed to develop a modularity measure to work on an unmodified version of the network (without aggregation or flattening). The traditional modularity measure was modified to match with a proper null model for multilayer networks. Having defined the new modularity measure, the authors embedded the new measure in traditional Girvan-Newman \cite{newman2004finding} and Louvain \cite{blondel2008fast} to detect multilayer networks communities. There is no need for modification of the network in this method; in turn, the method works on the multilayer networks which are directed or undirected. Modularity optimization methods have an inherent drawback, which is the resolution problem \cite{fortunato2007resolution}. \\

Interdonato et al. \cite{interdonato2017local} proposed a community detection method for multilayer networks that works on (un)directed and (un)weighted networks. The proposed algorithm is suitable for a wide range of networks, and the experimental result indicated its effectiveness on real-world datasets, such as heterogeneous or homogeneous networks. This algorithm is the first unsupervised multilayer community detection algorithm. In this method, three objective functions, which combine different inter-layer and intra-layer topological features, are optimized to extract a multilayer network's community structure. The optimization algorithm works iteratively. \\

\noindent \textbf{Metaheuristic Methods}\\
Community detection using modularity is an optimization approach. Metaheuristic methods aim to provide a good estimation for an optimization problem. Multi-objective optimization problems are popular methods for detecting the community structure of multilayer graphs. Genetic algorithms, as a metaheuristic, are widely used for optimizing uni-objective or multi-objective functions in order to extract the community structure of a multilayer network. In the followings, we elaborate on these metaheuristic methods.\\

Amelio et al. \cite{amelio2014cooperative} proposed a genetic algorithm for detecting communities in a multilayer network. The algorithm adopts a locus-based adjacency representation. In this representation, each row represents a layer of matrix, and each column indicates a node. Each node is initialized using a randomly selecting neighbor approach. Uniform crossover is used for modifying the chromosome. The fitness function of this algorithm is an extension of the traditional modularity measure. In this extension, the modularity measure of a pair of layers $Q_{i,j}$, for layer $i$ and $j$  is penalized if the detected community in the layer $i$ is not good in the layer $j$. Eventually, the modularity of the multilayer network is extracted from the pair-layer modularity measure of the multilayer network. For the nodes which stay out of the communities after finishing the genetic algorithm process, the communities are determined by a label propagation algorithm. Finally, a local search, based on the changes in the modularity measure after assigning a node to a community, is utilized to identify the best community of unassigned nodes. This algorithm can be used for unweighted networks which do not have any inter-layer relation.\\

Liu et al. \cite{liu2017improved} proposed an iterative optimization function that works separately on different layers and then calculates the similarity of the current layer with previous layers. This process ends up with a shared community structure. Modularity and NMI objective functions, along with the multi-objective genetic algorithm, are used in this method for uncovering community structure. The main aim of this method is maximizing the shared community structure and the modularity measure at the same time.\\

Similar to the \cite{liu2017improved}, Amelio et al. \cite{amelio2014uncovering} proposed a method that maximizes the modularity measure of the current layer concerning the previously considered layers' communities. This algorithm intends to maximize a multi-objective modularity optimization problem using a genetic algorithm. The algorithm tries to optimize the modularity measure, at the same time, minimize the difference of detected community in the current layer with the detected communities in the previously considered layers. The objective function of this method is a modularity measure, and the loss function NMI metric (Normalized Mutual Information), as the sharing cost.\\ 

In the first step of this algorithm, a simple genetic algorithm is employed to detect the community structure of the first layer. Next, a multi-objective genetic algorithm is used to identify communities in two stages. In the first stage, the modularity measure for the current layer is optimized, and in the second stage, NMI is optimized for the current layer and lastly considered layer. By doing so, the community structure result from the last layer is optimized to both of the objective functions. Thanks to using the metaheuristic algorithm, this algorithm is relatively fast and can be used on big networks.\\ 

There is an inborn drawback for these kinds of algorithms, which is inherited by genetic algorithm based multilayer community detection algorithms. This drawback is the problem of the local optimum detection, instead of global optimum detection, which makes the detected communities partially suitable. Amelio et al.'s study presents a framework in which other quality measures, instead of modularity, can be used, and by doing so, the resolution problem of modularity based methods can be eliminated. The input network of this algorithm should be undirected, and the nodes should be the same in all the layers. In fact, in \cite{amelio2014uncovering}, the algorithm tries to maximize the similarity of the community structure of the current layer with the very last considered layer. In this process, the impact of the first layers in the detected community is much less than the very last layers. The proposed method in \cite{ahmed2016multi} optimize a multi-objective function, in which the importance of the layers for detecting the community structure is the same.\\

MOGA-MDNet \cite{ahmed2016multi} uses a multi-objective genetic algorithm. The chromosome of the representation of this algorithm includes $n$ genes; each corresponds to a node in the network. In this representation, there is no need for a pre-specified number of communities. For detecting community structure, the algorithm starts with discovering the connected component. In this method number of the objective function is equal to the number of layers in the network. The genetic algorithm tried to find the best trade-off, which takes the community structure of all the layers into account simultaneously. This algorithm eliminates the intensive interaction impact on other dimensions. Additionally, the importance of a certain layer also can be considered in the algorithm. The proposed method in \cite{amelio2014community} aims to provide a new perspective to community detection problems by combining multi-objective genetic algorithms, local search, and the concept of temporal smoothness.\\

In \cite{amelio2014community} the authors describe the similarity of a multilayer network and a temporarily dynamic network, in which each layer corresponds to a network in a particular time step. Putting this together, they propose a new multi-objective genetic algorithm to revile the community structure of a multilayer network. First of all, the nodes in the first layer are clustered using a simple genetic algorithm with respect to modularity measure. Then, based on the node neighbors, the best cluster is selected for the nodes which are not assigned to any of the communities in the previous step. Afterward, a multi-objective genetic algorithm is employed on other layers with two objective functions correspond to modularity and NMI. If there are some unassigned nodes at the end of the algorithm, they are assigned to the communities based on their neighbors' clusters. This method can be used for heterogeneous multilayer networks.\\

MLMaOP is another multi-objective genetic algorithm that was proposed in \cite{pizzuti2017many}. This method utilizes NSGA-II \cite{deb2002fast} genetic algorithm for optimizing several objective functions. Having optimized the objective function, the algorithm uses one of the three strategies to select the best result from the Pareto front. Accurate clustering and the high-quality result are the main advantages of this method. However, the time complexity of this algorithm is high, and the algorithm quality reduces when the number of layer increase. The number of communities should also be specified beforehand, and this issue imposes confinement to this algorithm. Table \ref{tab_modu} summarize modularity-based methods proposed for multilayer community detection.\\

\begin{table*}[!]
	\begin{center}
		\caption{Table of modularity-based multilayer community detection methods.}
		\label{tab_modu}
		\begin{tabular}{|c|c|c|c|c|c|}
			\hline
			Reference&Algorithm Type& Weighted &Directed & Overlap & Evaluation Methods \\
			\hline
			\hline
			\cite{paul2016null}& Heuristic& -&- & - & \begin{tabular}{@{}c@{}}NMI\end{tabular}  \\
			\hline
			\cite{interdonato2017local}&Heuristic-Iterative&- & - & + & \begin{tabular}{@{}c@{}}Jaccard \\ Cosine\\ 3-cliques\end{tabular}  \\
			\hline
			\cite{song2015modularity}&Heuristic-GN& + & - & - & \begin{tabular}{@{}c@{}}NMI\\Accuracy \end{tabular}  \\
			\hline
			\cite{amelio2014cooperative}&Metaheuristic& - & - & - & \begin{tabular}{@{}c@{}}NMI \end{tabular}  \\
			\hline
			\cite{amelio2014uncovering}&Metaheuristic& + & + & - & \begin{tabular}{@{}c@{}}NMI \end{tabular}  \\
			\hline
			\cite{amelio2014community}&Metaheuristic& - & - & - & \begin{tabular}{@{}c@{}}NMI \\ Cross-dimension \end{tabular}  \\
			\hline
			\cite{ahmed2016multi}&Metaheuristic& - & - & - & \begin{tabular}{@{}c@{}}PMM \\ NMI \end{tabular}  \\
			\hline
			\cite{liu2017improved}&Metaheuristic& - & - & - & \begin{tabular}{@{}c@{}}NMI \end{tabular}  \\
			\hline
			\cite{pizzuti2017many}&Metaheuristic& - & + & - & \begin{tabular}{@{}c@{}}NMI \end{tabular}  \\
			\hline
		\end{tabular}
	\end{center}
\end{table*}

\subsubsection{Graph Feature Based Methods}
Some of the methods exploit network measures to uncover the community structure of multilayer networks. Having introduced a new clustering measure, known as the cross-layer edge clustering coefficient (CLECC), Brodka et al. \cite{brodka2011introduction}, introduced a new multilayer community detection methods which utilize this new measure. CLECC considers neighbors of a pair of nodes for calculating the strength of two nodes relation. In a hierarchical manner, similar to \cite{radicchi2004defining,girvan2001community}, in each step, the method compute CLECC of all the links and reduce the connection with minimum CLECC. These steps continue until the removal of the last link. CLECC can be used on multilayer networks, and this makes it different from \cite{radicchi2004defining,girvan2001community} which is only applicable to uni-layer networks.\\

Hmimida et al. \cite{hmimida2015community} extended the LICOD algorithm \cite{yakoubi2014licod}, which was proposed to detect community structure in uni-layer networks. This extended version works on multiplex networks. In the first step, the algorithm spots centric nodes. These nodes are ranked in descending order, with respect to their node degree in the multilayer network (the authors suggested a new definition for node degree in multilayer). A number of high ranked nodes are selected, and their Jaccard similarity is calculated. These similarities are used to put similar nodes in the same cluster. Then using the shortest path to the centric nodes, a new membership measure is calculated for each node whose cluster is not identified yet. Concerning this membership measure, the cluster of each node is identified. This method is simple to implement and understand; however, the time complexity of detecting the shortest path makes it impractical for big networks. Network feature-based methods are summarized in Table \ref{tab_net}.

\begin{table*}[!]
	\begin{center}
		\caption{Table of graph feature based multilayer community detection methods.}
		\label{tab_net}
		\begin{tabular}{|c|c|c|c|c|c|}
			\hline
			Reference&Algorithm Type& Weighted &Directed & Overlap & Evaluation Methods \\
			\hline
			\hline
			\cite{brodka2011introduction}&Network features& -&+ & -& \begin{tabular}{@{}c@{}}CLECC\end{tabular} \\
			\hline
			\cite{hmimida2015community}& Network features& -&+ & + & \begin{tabular}{@{}c@{}}Redundancy\end{tabular}  \\
			\hline
		\end{tabular}
	\end{center}
\end{table*}

\section{Evaluation Methods}
\label{sec_eval}
We discuss well-known evaluation methods that have been frequently used in the studies. Evaluation methods for uni-layer and multilayer networks can be divide into two main classes: quality-based evaluation measures and similarity-based evaluation measures. In the quality based measures, the quality of detected communities by an algorithm is compared with the quality of the detected communities by other methods. Indeed, there is no ground-truth label for evaluation data in this class. Similarity-based measures, mostly adopted from data mining, contrarily compare the detected community with the ground truth dataset. In the rest of this section, we describe the most popular evaluation methods.

\subsection{Quality-based Evaluation Methods}
Quality-based methods intend to assess the quality of the extracted community in the absence of ground truth data. To this end, these measures evaluate the ability of algorithms in detecting dense clusters or clusters in which the nodes inside a cluster are much more similar to nodes in other clusters. We introduce these measures below.\\

\textbf{Modularity for multilayer networks: }There are different modularity definition which stems from single-layer network modularity to be used in multilayer networks. We describe a version of these definitions, which was proposed \cite{zhang2017modularity}. We have the following equation for modularity in single-layer networks
\begin{align*}
\mathcal{H}(g) = &- \sum_{i \neq j} a_{ij} \underbrace{A_{ij} \delta(g_i,g_j)}_\text{Interanl existing edges}\\ 
&+ \sum_{i \neq j} b_{ij} \underbrace{(1 - A_{ij})\delta(g_i,g_j)}_\text{Internal non-existing edges} \\ \nonumber
&+ \sum_{i \neq j} c_{ij} \underbrace{A_{ij} [1 - \delta(g_i,g_j)]}_\text{External exiting edges} \\ \nonumber
&- \sum_{i \neq j} d_{ij} \underbrace{(1 - A_{ij})[1 - \delta(g_i,g_j)]}_\text{External non-esiting edges}, \\ \nonumber
\end{align*}
where $A_{ij}$ is the weights of edge between $i$ and $j$, $g_i$ demonstrates the label of community of node $i$, and $a$, $b$, $c$, and $d$ are free parameters. This equation can be expanded for multilayer networks as follows:
\begin{align*}
\mathcal{H_M} (g) = &- \sum_{i \neq j} \sum_{v = 1}^{F} \sum_{s = 1}{V_V} \underbrace{a_{ijs}^{\{v\}}A_{ijs}^{\{v\}}\delta(g_{is}^{\{v\}},g_{js}^{\{v\}})}_\text{Within-layer internal exiting edge} \\
&+ \sum_{i \neq j} \sum_{v = 1}^{F} \sum_{s = 1}{V_V} \underbrace{b_{ijs}^{\{v\}}(1-A_{ijs}^{\{v\}})\delta(g_{is}^{\{v\}},g_{js}^{\{v\}})}_\text{Withing-layer internal non-exiting links}  \\ \nonumber
&+ \sum_{i \neq j} \sum_{v = 1}^{F} \sum_{s = 1}{V_V}  \underbrace{c_{ijs}^{\{v\}}A_{ijs}^{\{v\}}\big[1-\delta(g_{is}^{\{v\}},g_{js}^{\{v\}})\big]}_\text{Within-layer external exiting links} \\ \nonumber
&- \sum_{i \neq j} \sum_{v = 1}^{F} \sum_{s = 1}{V_V}  \underbrace{d_{ijs}^{\{v\}}\big(1 -A_{ijs}^{\{v\}}\big)\big[1-\delta(g_{is}^{\{v\}},g_{js}^{\{v\}})\big]}_\text{Within-layer external non-exiting links} \\ \nonumber
&- \sum_{sw \neq rw} \sum_{i = 1}^{N} \underbrace{e_{isr}^{\{vw\}}C_{isr}^{\{vw\}}\delta\big(g_{is}^{\{v\}}, g_{ir}^{\{w\}}\big)}_\text{Between-layer internal exiting coupling} \\ \nonumber
&+ \sum_{sw \neq rw} \sum_{i = 1}^{N}  \underbrace{f_{isr}^{\{vw\}}\big(1 - C_{isr}^{\{vw\}}\big)\delta\big(g_{is}^{\{v\}}, g_{ir}^{\{w\}}\big)}_\text{Between-layer internal non-exiting coupling} \\ \nonumber
&+ \sum_{sw \neq rw} \sum_{i = 1}^{N}  \underbrace{g_{isr}^{\{vw\}}C_{isr}^{\{vw\}}\big[1-\delta\big(g_{is}^{\{v\}}, g_{ir}^{\{w\}}\big)\big]}_\text{Between-layer external exiting couplings} \\ \nonumber
&- \sum_{sw \neq rw} \sum_{i = 1}^{N} \underbrace{h_{isr}^{\{vw\}}\big(1-C_{isr}^{\{vw\}}\big)\big[1-\delta\big(g_{is}^{\{v\}}, g_{ir}^{\{w\}}\big)\big]}_\text{Between-layer external non-exiting couplings}, \\ \nonumber
\end{align*}
where $s$ and $r$ indicate different layers, $v$, and $w$ stands for the the aspects. In this formula, $N$ and $V_V$ demonstrate total number of nodes in a layer  and total number of layers of aspect $v$. Matrix $A$, $C$, and g represents the within-layer adjacency, between-layer adjacency, and community label matrix, respectively. $a_{ijs}$, $b_{ijs}$, $c_{ijs}$, $d_{ijs}$, $e_{ijs}$, $f_{ijs}$, $g_{ijs}$, and $h_{ijs}$ are free parameter to adjust the formula.\\

\textbf{Cross-Layer Edge Clustering Coefficient (CLECC): }CLECC \cite{brodka2011introduction} is indeed edge version of clustering coefficient. This measure denotes how much the neighbors of the user $x$ and neighbors of the user $y$ are interconnected, and is defined as follows:
\begin{equation}
ECC(x,y) = \frac{z_{x,y} + 1}{S_{x,y}},
\end{equation}
where $x$ and $y$ are the nodes which are connected by link $<x,y>$, $z_{x,y}$ is the number of 3-cliques made on the edge $<x,y>$ and all edges between $x$, $y$, and their neighbors, and $S_{x,y}$ is the possible number of 3-cliques that can be built based on edge $<x,y>$ and all possible edges.\\

\textbf{Homophily: } Homophily is defined in \cite{alimadadi2019community} and denotes the diversity of edges between nodes, and reflect to what extent there are some edges in the different layer between a pair of nodes. The formula of homophily is as follows:
\begin{equation*}
r = \frac{\sum_i e_{ii} - \sum_i a_ib_i}{1 - \sum_i a_i b_i},
\end{equation*}
where 
\begin{equation*}
a_i = \sum_j = e_ij,
\end{equation*}
\begin{equation*}
b_j = \sum_i e_{ij},
\end{equation*}
and $e_{ij}$ is the link between $i$ and $j$.\\

\textbf{Redundancy: }Redundancy \cite{hmimida2015community} demonstrates to what extent the link between a pair of nodes also occurs in other layers. This measure is similar to the homophily measure but somehow different. If we have
\begin{itemize}
	\item $P$ is the set of couple $(u,v)$, which are directly connected to at least one layer.
	\item $\bar{P}$ is the set of couple $(u,v)$, which are directly connected in at least two layers.
	\item $P_c \subset P$ represents all links in the community $c$.
	\item $\bar{P}_c \subset \bar{P}$ is the subset of $\bar{P}$, and which are also in $c$. 
	\item $\alpha$ is the number of layers.
\end{itemize}
Then the redundancy can be calculated as follows:
\begin{equation}
\rho (c) = \sum_{(u,v) \in \bar{P}_c} \frac{||\{k: \exists A_{uv}^{[k]} \} \neq 0\}||}{\alpha \times ||P_c||}.
\end{equation}
Eventually, the quality of a partitioning is defined as follows:
\begin{equation*}
\rho(\mathcal{P}) = \frac{1}{||\mathcal{P}||} \sum_{c \in \mathcal{P}} = \rho(c)
\end{equation*}
\subsection{Similarity based}
The other group of evaluation methods mostly stem from data mining. In the classification problem of data mining, we have a set of label data. The algorithms aim to use part of the data for developing a prediction model and use the rest of the data for evaluating to what extent the predicted method is working well. Some of the measures which have been used in data mining and information retrieval are adopted by the community detection research field to assess the proposed data. In this evaluation approach, the labeled dataset is required.\\

\textbf{Jaccard similarity:} Jaccard is a statistical measure for calculating the similarity of two sets. Although this measure is interpretable, it is not robust, and a small loss in data can end up with a significant change in the Jaccard similarity result. Jaccard similarity formula for multilayer network is as follows:

\begin{equation*}
Jaccard\_sim_{i,j} (u,v) = \frac{|N_i(u) \cap N_j(v)|}{|N_i(u) \cup N_j(v)|},
\end{equation*}
where $N_i(u) = \{v \in V | (v,u) \in E_i\}$ indicates the set of neighbors of a node $u$ in layer $i$, $V$ is the set of nodes, and $E_i$ is the set of edges in layer $i$.\\

\textbf{Cosine similarity: }Cosine similarity is another famous similarity measure that calculates the angle of two vectors, and this angle indicates how much two vectors are similar to each other. For a multilayer network, in which $i$ and $j$ are two layers, the cosine formula is as follows:

\begin{equation*}
Cosine\_sim = \frac{|N_i(u) \cap N_j(v)|}{\sqrt{|N_i(u)||N_j(v)|}},
\end{equation*}
where, again, $N_i(u) = \{v \in V | (v,u) \in E_i\}$ indicates the set of neighbors of a node $u$ in layer $i$, $V$ is the set of nodes, and $E_i$ is the set of edges in layer $i$.\\

\textbf{Triad-based similarity:} Jaccard similarity, calculated on the set of 3 cliques (complete graph with three nodes), is called triad-based similarity. If $T_i(u)$ demonstrates the set of 3-cliques which node u belongs to in layer $i$, we have:
\begin{equation*}
Triad\_sim_i,j(u,v) = \frac{|N_i(u) \cap N_j(v)|}{|T_i(u) \cup T_j(v)|},
\end{equation*}
where, $N_i(u) = \{v \in V | (v,u) \in E_i\}$ indicates the set of neighbors of a node $u$ in layer $L_i$, $V$ is the set of nodes, and $E_i$ is the set of edges in layer $i$.\\

\begin{figure*}
	\centering
	\includegraphics[width=0.5\textwidth]{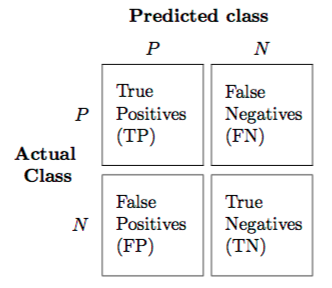}
	\caption{Matrix of possible results of a prediction.}
	\label{confu_pic}
\end{figure*}

\textbf{Precision: } Precision indicates the fraction of the result which are relevant. Consider Figure \ref{confu_pic}; we have the following formula for precision:
\begin{equation}
Precision = \frac{True~Positive}{True~Positive + False~Positive}
\end{equation}
\textbf{Recall: } Recall is the percentage of total relevant results correctly labeled by the algorithm. Considering Figure \ref{confu_pic} we have the followings formula for recall:
\begin{equation*}
Recall = \frac{True~Positive}{True~Positive+False~Negative}
\end{equation*}
\textbf{$F1$ score: } $F1$ score is harmonic average of \textit{Precision} and \textit{Recall}.

\textbf{Accuracy: }Considering Figure \ref{confu_pic} we have following formula for accuracy:

\begin{equation*}
Accuracy = \frac{TP}{TP + TN + FP + FN},
\end{equation*}
\textbf{NMI: }NMI (Normalized Mutual Information) is a measure for calculating the similarity of two clusters. Let's $P_{pred} = \{\hat{\pi}_1, \dots, \hat{\pi}_k\}$ is predicted communities and $P_{true} = \{{\pi}_1, \dots, {\pi}_l\}$ is true clustering of a network. We have $n_{ij} = |\pi_j \cap \hat{\pi}_i|$ for $j = 1, \dots, k$ and $i = 1, \dots, l$, the NMI is as followings:
\begin{equation*}
NMI = \frac{2 \sum_{i = 1}^{l}\sum_{j=1}^{k} n_{ij} \log\frac{n~n_ij}{n_i~\hat{n}_j}}{\sum_{j=1}^{k}\hat{n}_j \log\frac{\hat{n}_j}{n} + \sum_{i = 1}^{l} n_i \log \frac{n_i}{n}}
\end{equation*}
\textbf{Rand Index (RI): }RI is a measure which shows the similarity of two data clustering. Let's assume.
\begin{itemize}
	\item $a$, the number of the pairs of elements in $S$ that are in the same subset in $X$ and in the same subset in $Y$;
	\item $b$,  the number of the pairs of elements in $S$ that are in the different subsets in $X$ and in the different subsets in $Y$;
	\item $c$, the number of pairs of elements in $S$ that are in the same subset in $X$ and in the different subsets in $Y$;
	\item $d$, the number of pairs of elements in $S$ that are in the different subsets in $X$ and in the same subset in $Y$.
\end{itemize}
Then we have this formula for RI:
\begin{equation}
RI = \frac{a + b}{a + b + c + d}
\end{equation}
\textbf{Adjusted Rand Index(ADI): }ADI is another form of RI and indicates the chance of elements grouping. If for a set $S$ of $n$ elements two detected clusterings is $X = \{X_1, X_2, \dots, X_r\}$ and $Y=\{Y_1,Y_2, \dots, Y_s\}$, and $n_{ij} = |X_i \cap Y_j|$ denotes the number of objective common between $X_i$ and $X_j$. Then we have the following formula for ADI:

\begin{equation}
ARI = \frac{\sum_{ij}\binom{n_{ij}}{2} - \big[\sum_i \binom{a_i}{2}\sum_j \binom{b_j}{2}\big] \big/ \binom{n}{2}}{\frac{1}{2}\big[\sum_i \binom{a_i}{2} + \sum_j \binom{b_j}{2}\big] - \big[\sum_i \binom{a_i}{2} + \sum_j \binom{b_j}{2}\big] \big/ \binom{n}{2}}
\end{equation}  
\textbf{Purity: }Purity calculates the accuracy of clustering by counting the number of correctly assigned nodes to the clusters divided by the number of nodes $N$.
\begin{equation*}
purity(\Omega,C) = \frac{1}{N} \sum_k max_j |\omega_k \cap c_j|,
\end{equation*}
where $\Omega = \{\omega_1,\omega_2,\dots, \omega_k\}$ is the detected communities and $C = \{c_1, c_2, \dots, c_J\}$ is the true communities.

\section{Challenges}
\label{sec_chall}
This section is dedicated to comparing the methods, discussing the approaches pros and pons, and finally introducing challenges and future road map.\\

\subsection{Methods Comparison}
As mentioned before, indirect methods for multilayer community detection are divided into two main groups, namely, flattening and ensembling. There are some benefits to using the indirect method. One of the most important benefits is the simplicity of the method. Directed techniques need to consider the information of all layers for uncovering the community structure of a multilayer network, which this issue makes direct methods difficult to understand and implement. Another advantage of indirect methods is their time complexity, which often is less than direct methods because uni-layer network community detection is a well-studied topic, and the proposed methods are efficient enough. However, multilayer community detection research field is in its infancy, and it needs a significant amount of efforts.\\

Another specific advantage of ensembling methods is that for each layer, a particular method of community detection can be employed, then the multilayer community structure is extracted using a form of consensus. This attribute provides us sufficient flexibility to adopt the best possible community detection, such as, attributed, directed, or weighted, to satisfy each layer's specific requirements.\\

\begin{figure*}
	\centering
	\includegraphics[width=0.5\textwidth]{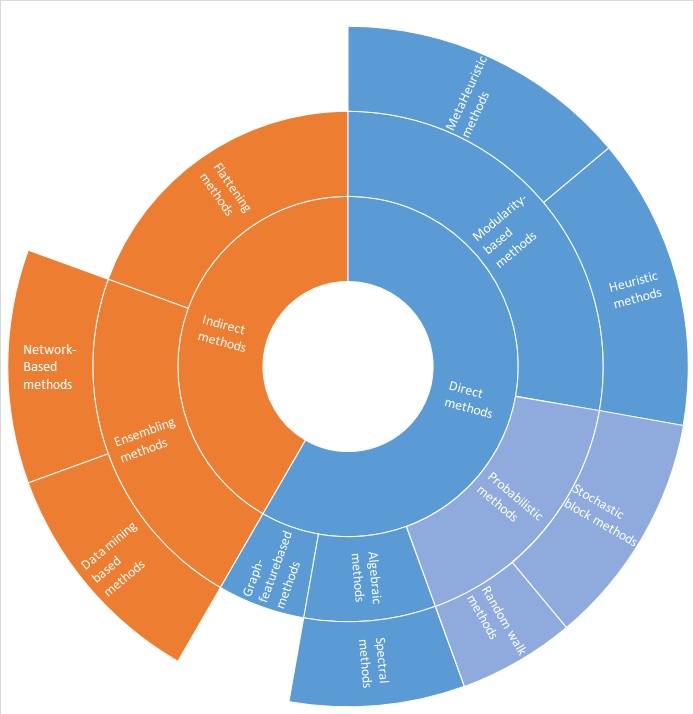}
	\caption{Distribution of multilayer community detection studies.}
	\label{overall_pic}
\end{figure*}

On the other hand, the main disadvantage of indirect methods is neglecting inter-layer edges, which means putting a considerable amount of information out of calculations. Another significant drawback of indirect methods is that these methods do not take into account all the information in the dataset. Unlike direct methods, the flattening methods compress all information of the multilayer network into a single layer, therefore, loss a significant amount of valuable information, and this lost information can not be retrieved down the road. In the ensembling methods, first, the community structures of each layer are revealed, then using a consensus method, the most frequent communities are returned as the final results. Because the direct methods use the complete network, unlike the indirect methods, they are much more interpretable. Figure \ref{overall_pic} indicates that direct methods have attracted more attention compared to indirect methods. Among direct approach modularity-based methods are the most popular followed by probabilistic methods.\\

Directed multilayer community detection methods are more complex than undirected multilayer community detection methods; thus, most of the proposed methods only deal with indirect networks. Only five of the considered methods are capable of working on direct networks; whereas, the indirect methods utilize any proposed uni-layer methods for community detection in each layer, and community detection in directed uni-layer networks are broadly studied in the past few years. Therefore, indirect methods are more capable of handling direct networks. Among the studied indirect methods, 10 of them can work on directed graphs.\\

Another persistent challenge in community detection, both uni-layer and multilayer networks, is detecting overlap communities. Among the studied methods just three modularity-based algorithms and three probabilistic methods can reveal overlapped community structure.\\

\subsection{Future Challenges}
As mentioned above, multilayer community detection is in its infancy; therefore, we believe there is a long way in front of the researchers. In this section, we discuss current and future challenges of this popular research field.\\

\subsubsection{Lack of Standard Definitions}
The very first step in a research field is making an agreement on the definition of the essential concepts. There is almost no widely agreed definition in the multilayer community detection research field. Thus, the proposed definitions are far from each other. This issue leads to various algorithms that are presented from different perspectives. It seems much more effort is needed for making a more strong background for this research field. Indeed, providing standard definitions and agreeing on it is the first step for shedding light on the future road map of this research field.

\subsubsection{Standard Dataset}
\begin{figure*}
	\centering
	\begin{subfigure}[t]{0.4\textwidth}
		\includegraphics[width=\textwidth]{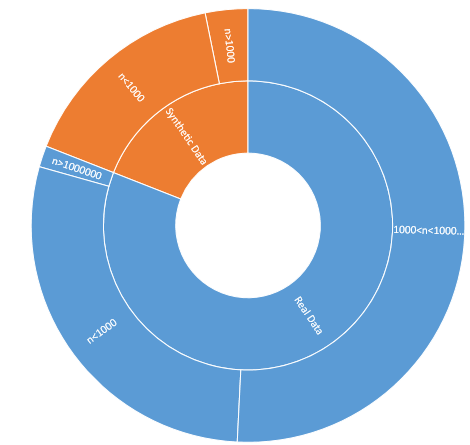}
		\caption{Degree and layer distribution of the datasets.}
		\label{degree_pic}
	\end{subfigure}\hspace{15mm}
	\begin{subfigure}[t]{0.4\textwidth}
		\includegraphics[width=\textwidth]{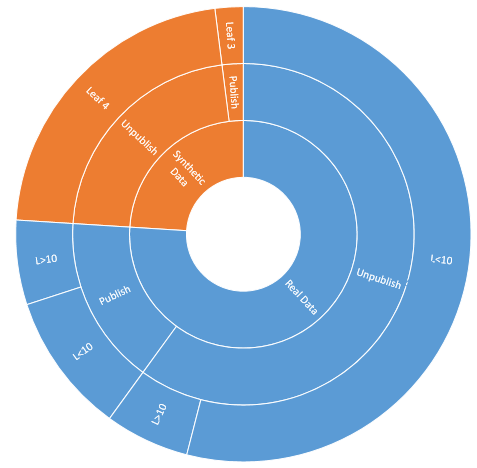}
		\caption{Availability of the datasets.}
		\label{publish_pic}
	\end{subfigure}
\end{figure*}

There are a few publicly available datasets, which this shortage makes evaluation of the proposed method difficult. Additionally, there is no standard benchmark for multilayer networks to be used for assessment. Each of the proposed algorithms has been evaluated on different datasets. Therefore, the comparative study of the proposed methods is impossible. In the lack of standard benchmarks, no comprehensive comparative studies have been done on the multilayer networks. Hence, we believe providing a standard benchmark and then doing a comparative study on the proposed method will pave the way for future studies by reviling the shortcomings of current works. We divide the used datasets in the studies into two groups, that is, synthetic and real-world datasets. Figure \ref{degree_pic} and \ref{publish_pic} indicates the features of different kinds of datasets. As it is evident, most of the real-world datasets are not publicly available (Figure \ref{publish_pic}). Figure \ref{degree_pic} denotes the degree and layer distribution of the networks. It is clear that the number of layers for most of the datasets is fewer than ten. Hence, more complex datasets are required for more realistic evaluations of the networks.

\subsection{Identifying Layers' Dependencies}
Detecting layers' dependencies can enhance the quality of detected communities remarkably because a significant amount of information lies on these dependencies. To determine layer dependencies, we need new definitions and novel measures for calculating the dependencies. There is a small amount of effort allocated to resolving multilayer network dependencies. Similarly, among the studied algorithms, there is no multilayer community detection method that considers the layers' dependencies. It seems, in down the road more attention will be attracted to this subject.

\subsubsection{Scalability}
In the age of the interactive web, the amount of information has increased significantly; therefore, algorithms with less time and memory complexity is necessary to deal with the enormous amount of data which is available nowadays. Although new attempts are dedicated to increasing the scalability of the proposed method, most of the proposed methods are not able to handle big data. Other methods to cope with big data is a realistic sampling of multilayer networks approach for reducing the network size. In our opinion, more research will go over this topic in the future.

\subsubsection{Temporal Analysis}
Dynamic networks have attracted research attention in the past few years. Taking into account the temporal dimension of networks will provide more in-depth insight into the multilayer networks. Though a number of studies have done on the temporal dimension of uni-layer networks \cite{he2015fast}, this aspect of networks is mostly neglected in multilayer networks; Bazzi et al. \'s work \cite{bazzi2016community} is one of the a few works, aimed to combine the temporal aspect into multilayer community detection. Considering the temporal aspect of multilayer networks adds an incredible amount of complexity to the topic, which requires more studies.

\subsubsection{Evaluation}
There is an emergent need for a meaningful evaluation framework for multilayer networks due to the lack of standard evaluation methods. Most of the utilized algorithms have used uni-layer evaluation methods, which are not able to evaluate all the facets. This is another line of future studies that can enrich the multilayer community detection research field.

\section{Conclusion}
\label{sec_concl}
In this study, we reviewed the proposed algorithm for multilayer community detection. Two primer line of approaches is detectable among the methods: direct and indirect methods. Indirect methods are flattening or ensembling algorithms. For direct methods, we divided the proposed algorithms into four main groups, namely, algebraic, probabilistic, modularity-based, and graph feature-based. We deeply elaborate on proposed methods.\\

Additionally, we comprehensively explained the pros and cons of each method. Then, we discussed the challenges and future works. The most important challenges, which we believe will be tackled in the near future, are the lack of standard definition and shortage in the available dataset for evaluation. More theoretical studies are necessary for making a strong background for the multilayer community detection research field. Also, more efforts should be dedicated to directed multilayer networks in the future. Extracting overlapping community structure is another interesting topic, which the previous study only scratched its surface. Proposing a standard framework will provide standard environment for future research.



\bibliographystyle{unsrt}  

\bibliography{cas-refs}

\end{document}